\documentclass[conference]{IEEEtran}
\usepackage{times}
\usepackage[lined,algonl,ruled]{algorithm2e}
\usepackage{algpseudocode} \usepackage{float} \usepackage{graphicx}
\usepackage{amssymb} \usepackage{comment} \usepackage{caption}
\usepackage{subcaption} \usepackage{multirow} \usepackage{longtable}
\usepackage[hyphens]{url} 

\title{Title}
\author{
    \IEEEauthorblockN{Muyuan
      Li\IEEEauthorrefmark{1}\IEEEauthorrefmark{2},
      Haojin Zhu\IEEEauthorrefmark{1}, Zhaoyu
      Gao\IEEEauthorrefmark{1},
      Si Chen\IEEEauthorrefmark{2},
      Kui Ren\IEEEauthorrefmark{2},
      Le Yu\IEEEauthorrefmark{1},
      Shangqian Hu\IEEEauthorrefmark{1}
    }
    \IEEEauthorblockA{\IEEEauthorrefmark{1}Department of Computer
      Science and Engineering\\
      Shanghai Jiao Tong University\\
      \{leilmyxwz, zhuhaojin, gaozy1987, lewisyu24, hu.shangq\}@gmail.com}
    \IEEEauthorblockA{\IEEEauthorrefmark{2}Department of Computer
      Science and Engineering\\
      University at Buffalo\\
      \{schen23, kuiren\}@buffalo.edu}
}

\begin{document}
\title{All Your Location are Belong to Us: Breaking Mobile Social
  Networks for Automated User Location Tracking}

\maketitle
\thispagestyle{empty}

\begin{abstract}
  Many popular location-based social networks (LBSNs) support built-in
  location-based social discovery with hundreds of millions of users
  around the world. While user (near) realtime geographical
  information is essential to enable location-based social discovery
  in LBSNs, the importance of user location privacy has also been
  recognized by leading real-world LBSNs. To protect user's exact
  geographical location from being exposed, a number of location
  protection approaches have been adopted by the industry so that only
  relative location information are publicly disclosed. These
  techniques are assumed to be secure and are exercised on the daily
  base. In this paper, we question the safety of these
  location-obfuscation techniques used by existing LBSNs.  We show,
  for the first time, through real world attacks that they can all be
  easily destroyed by an attacker with the capability of no more than a
  regular LBSN user. In particular, by manipulating location
  information fed to LBSN client app, an ill-intended regular user can
  easily deduce the exact location information by running LBSN apps as
  location oracle and performing a series of attacking strategies.
  We develop an automated user location tracking system and test it on
  the most popular LBSNs including Wechat, Skout and Momo. We
  demonstrate its effectiveness and efficiency via a 3 week real-world
  experiment with 30 volunteers. Our evaluation results show that we
  could geo-locate a target with high accuracy and can readily recover
  users' Top 5 locations. We also propose to use grid reference system
  and location classification to mitigate the attacks. Our work shows
  that the current industrial best practices on user location privacy
  protection are completely broken, and it is critical to address this
  immediate threat.
\end{abstract}

\section{Introduction}
Mobile social networks have gained tremendous momentum since recent
years due to both the wide proliferation of mobile devices such as
smartphones and tablets as well as the ubiquitous availability of
network services. Millions of users are enabled to access and interact
with each other over online social networks via their mobile
devices. Moreover, the positioning technologies such as GPS, and
wireless localization techniques for mobile devices have made both the
generation and sharing of real-time user location updates readily
available. This, in turn, leads to the extreme popularity of
location-based social networks (LBSNs) such as Facebook Places, Google
Latitude, PCube, Foursquare, Wechat, Momo, Badoo, Grindr, Blendr, and
Tapmee, which boost up to hundreds of millions of users. As one of the
most popular LBSNs in China, Wechat achieved more than 300 million
registered user accounts in only two years, and is used in over 200
countries \cite{whats-app}. Another LBSN app Momo has 30 million
users, 2.2 million of whom use the app on a daily base
\cite{smith13:_how_many_people_use_top, ifeng:momo-announce}. Skout, a
very popular dating app in North America, draws 1.5 million new users
a month who check into the app an average of nine times a day
\cite{apps13:_skout}.

In contrast to traditional LBSNs such as Foursquare, which allow users
to check-in at locations and share the information with friends within
vicinity, the newer ones put tremendous focuses on location-based
social discovery as the latest trend. Location-based social discovery
explicitly enables on-the-spot connection establishments among users
based on their physical proximity. Examples of such
LBSNs include Google Latitude, PCube, Wechat, Momo and Skout. While
services like Google Latitude and PCube allow their users to control
with whom they want to share the location information, popular ones
like Wechat, Momo and Skout allow location-based social discovery
solely based on users' physical proximity.

Along with the popularity of location-based social discovery is the
increasing danger of user privacy breaches due to location information
exposure. Recent studies have shown that four spatiotemporal points
are sufficient to uniquely identify the individuals in an anonymized
mobility data set \cite{9-de2013unique,
  10-Ma:2010:PVP:1859995.1860017} and little outside or social network
information is needed to re-identify a targeted individual or even
discover real identities of users \cite{9-de2013unique,
  11-srivatsa2012deanonymizing,
  12-zang2011anonymization}. Furthermore, users' location traces can
leak much information about the individuals' habits, interests,
activities, and relationships as pointed out in
\cite{6-schilit2003wireless}. And loss of location privacy can expose
users to unwanted advertisement and location-based spams/scams, cause
social reputation or economic damage, and make them victims of
blackmail or even physical violence.

Recognizing the danger of user location privacy leakage due to the use of mobile device in general,
various research efforts have been
devoted to location privacy. Most of them focus on developing the
general location privacy protection mechanisms for location-based
services (LBSs) that allow users to make use of LBSs while limiting
the amount of disclosed sensitive information
\cite{18-bindschaedler2012track, 19-beresford2003location,
  20-freudiger2009optimal, 21-meyerowitz2009hiding,
  22-shokri2012protecting, 23-xu2009feeling, 24-shokri2011quantifying,
  25-gedik2005location}. Example techniques include anonymous service
uses, cloaking based technique\cite{23-xu2009feeling}, mixzone or
silent period\cite{20-freudiger2009optimal,
  24-shokri2011quantifying}. Mechanisms are also proposed to enable
proximity testing without revealing the mobile users' real location
information \cite{16-zheng2012sharp, 17-narayanan2011location} for
privacy preserving distributed social discovery.

User location privacy in real-world LBSN apps, however, has not
received enough attention. Current industrial practices are yet to be
scrutinized for their (in)adequacy, and users are usually kept in the
dark for the potential risks they face. Different from directly access
(e.g., iAround, SayHi) or authorized access type LBSNs (e.g., Google
Latitude and PCube), some popular apps including Wechat, Momo, and
Skout hide the exact location of mobile users by only sharing the
relative distances among the users, limiting the localization accuracy
to a certain range or restricting the display coverage to a particular
area.

These location obfuscation techniques are expected to enable
location-based social discovery, while at the same time protect
users' location privacy. And millions of users are made to believe so
and thus fail to be conscious about the potential risk of leaking
their location information when using the services. This could be
potentially more dangerous than the former case of exact location
exposure as in Banjo, etc., should these hiding techniques
fail, because in the former case users are explicitly aware of the
risk and can thus either proactively avoid it by logging off the
service when necessary or voluntarily take it.

In this paper, we ask and answer two fundamental questions regarding
user privacy in the most popular LBSNs protected by the-state-of-art
location hiding techniques. First, is it possible to make an
involuntary localization of a random LBSN user by exploiting the
public available information only? That is, without hacking into the
services and using only the client side information that publicly
available through the unmodified app of LBSNs, could we accurately
localize a random online user of no priori knowledge? Second, could we
freely track a particular user within a reasonably short time period?
By investigating three most popular LBSN apps (Wechat, Momo and
Skout), our answers to these two questions are more than a simple
``yes''. Our research findings show that: 1) An attacker could perform
a range-free, involuntary user localization with high localization
accuracy; 2) Furthermore, it can successfully establish very accurate
user location traces.

We implement FreeTrack, an automated user location tracking system for
mobile social networks, which could automatically track Wechat, Skout
and Momo even without users' awareness. To demonstrate the
effectiveness of FreeTrack, we perform a three-week real-world attack
towards 30 volunteers from United States, China and Japan. By
comparing the collected users' real trace with the inferred trace, it
is found that the mean tracking error of FreeTrack is 51m for 74
Wechat tests, 25m for 119 Momo tests, and 130m for 156 Skout tests.
What's more, users' Top 5 locations could be easily identified by the
attacker.  According to the existing works, more than $50\%$ of the
individuals could be uniquely identified if given top 2
locations\cite{9-de2013unique}. Hence, the newly identified
attacks pose a
serious threat towards the locations privacy of hundreds of millions
of LBSN users.

The rest of paper is organized as follows: Section \ref{sec:lbsns} is
the classification of LBSNs. Section \ref{sec:method} describes our
attack methodology, which is followed by Section
\ref{sec:implementation} describing the implementation of the
attack. Section \ref{sec:evaluation} presents the evaluation results.
In Section \ref{sec:mitigation}, the mitigation approaches are
discussed and Section \ref{sec:related} summarizes the related
work. Finally, Section \ref{sec:conclusion} concludes the paper.


\begin{table*}[t]
  \centering
  \small
  \begin{tabular}{l l l l l l l l l l}
    \hline
    & Distance & Accuracy Limit & Coverage Limit & Number of Users  & Platform & SDK & Category\\
    &         &   &   & (millions) & or region  &  &  &\\
    \hline
    Wechat & Y & 100m & 1km (shanghai) &  300 millions & iOS/Android/WP & Google & II\\
    Skout  & Y & 0.5mile & N/A &  5 millions & iOS/Android/WP & Google & II\\
    Momo   & Y & 10m  & N/A &  30 millions & iOS/Android/WP & Baidu & II\\
    Whoshere & Y & 100m & N/A &  5 millions in 2012 &  iOS/Android & Google & II\\
    MiTalk & Y & 100m & 0.6km (shanghai)  & 20 millions & iOS/Android & Baidu & II\\
    Weibo & Y & 100m & 1600m & 500 millions & iOS/Android/WP & Google & II\\
    SayHi & Y & 10m & 1000km & 500 thousands &iOS/Android & Google & I/II \\
    iAround & Y & 10m & N/A & 10 millions & iOS/Android & Baidu  & I/II \\
    Duimian & Y & 100m & N/A & 500 thousands & iOS/Android & Google & II\\
    Doudou Friend & Y & 10m & N/A & 1 million & iOS/Android & Amap & II\\
    $U+$ & Y & 10m & N/A & 10 millions & iOS/Android & Baidu & II\\
    Topface & Y & 100m & N/A & 50 million & iOS/Android & Google & II\\
    Niupai & Y & 10m & N/A & 61 thousands & iOS/Android & Google & II\\
    LOVOO & Y & 100m & 27.8km (shanghai) & & iOS/Android & Google & II\\
    KKtalk & Y & 10m & N/A & 320 thousands &  iOS/Android & Google &
    II\\
    Meet24 & Y & 0.5mile & N/A & &  iOS/Android & Google & II\\
    Anywhered & Y & 10m & N/A & 750 thousands &  Android & Baidu & II\\
    I Part & Y & 10m & 1000m & 8 millions &  iOS/Android & Google
    &II\\
    Path & N & N/A & N/A & 10 millions & iOS/Android & Google & I \\
    TweetCaster & N & N/A & N/A & 10 millions &  iOS/Android/WP &
    Google & I\\
    Google Latitude & N & N/A & N/A & 10 millions &
    iOS/Android/WP & Google & I\\
    eHarmony & N & N/A & N/A & 5 millions & iOS/Android & Google & I\\
    SinglesAroundMe & N & N/A & N/A & 1 million & iOS/Android & Google & I\\
    \hline
  \end{tabular}
  \caption{Location based friend discovery apps}
  \label{tbl:apps}
\end{table*}

\section{LBSN: The State-of-the-Art}
\label{sec:lbsns}

\subsection{Classification of LBSNs}
With the wide use of mobile devices, and the increasing attention on
mobile social networking, location-based social networks (LBSN)
focusing on the small local social network derived from a user's
geographical location become increasingly popular. In addition to the
conventional location-based user check-in apps (e.g., Foursquare),
more LBSN apps are exploiting the users' geographical information to
achieve distance-based social discovery and location
sharing. Based on how real-world LBSNs share the location information
among their users in order to allow location-based social discovery,
they can be classified into two main categories: I) LBSNs with Exact
Location Sharing and II) LBSNs with Indirect Location Sharing. Table
\ref{tbl:apps} is a summary from our survey of 20 popular real-world LBSNs.

Category I has two subtypes. The Subtype I is \emph{Open Access
  Location Sharing}. These LBSNs present the exact locations of their
users without any restriction. Take Banjo as an example. By clicking
``Places'' tab, the users are allowed to see people of the same city,
the exact location of which are explicitly displayed on a
map. Similar applications include SayHi, I-Am, iAround. The
Subtype II is \emph{User Authorized Location Sharing}. For this type
of LBSNs, users can have the control to choose with whom they share
their exact location information. For example, in Google Latitude or
PCube, a user can define the set of other users (or friends) who are
allowed to see his position on the map. In general, when the users
choose this kind of apps, they should have a clear idea about the
potential location privacy risk and be willing to share their location
information with other LBSN users or only share their location with
their trusted friends.
%

In Category II, the exact geographic information is hidden or
obfuscated by a series of location privacy protection
techniques. Different from Category I which reveals users' exact
locations, in Category II LBSNs, users are assured that their exact
location information are never shared for the purpose of privacy
protection. To achieve this goal, LBSN service providers adopt the
following location obfuscating techniques.

\noindent\textbf{I. Relative Distance Only}: This has been a very common
location hiding technique adopted by many popular LBSNs, including Wechat,
Skout, and Momo. Users in this case can only see others' geographic
distances instead of location coordinates. From the user's point of
view, revealing the distances rather than coordinates could hide the
exact location but still allow the potential near strangers (or
potential friends) to discover the presence of this user.

\noindent{\textbf{II. Setting the Minimum Accuracy Limit}:} Setting a safe
localization accuracy limit is a traditional
location obfuscation technique \cite{ardagna2007location}.
Most of the LBSN apps predefine a certain minimum accuracy limit for
geo-localization to further protect the users' exact location.
For
example, Skout defines localization accuracy to 1 mile, which means
that the users will be located with an accuracy no better than 1
mile. Similarly, Wechat and Momo set 100m and 10m as their
localization accuracy limits.

\noindent{\textbf{III. Setting the Localization Coverage Limits}:} To prevent
malicious users from abusing the geo-localization, an additional
functionality, Localization Coverage limit is provided to restrict the
users' localization capability to a specific region or under the
maximum number of displayed users. For example, Wechat only displays
the relative distance of users, the number of which is less than a
predefined threshold (e.g., 800m in wechat for a high user density
region).

In addition to above mentioned location hiding techniques, there are
other factors which contribute to the localization errors, which will be presented as follows.

\begin{figure*}[htbp!]
  \centering
  \begin{subfigure}[htbp!]{0.3\textwidth}
    \centering
    \includegraphics[width=\textwidth]{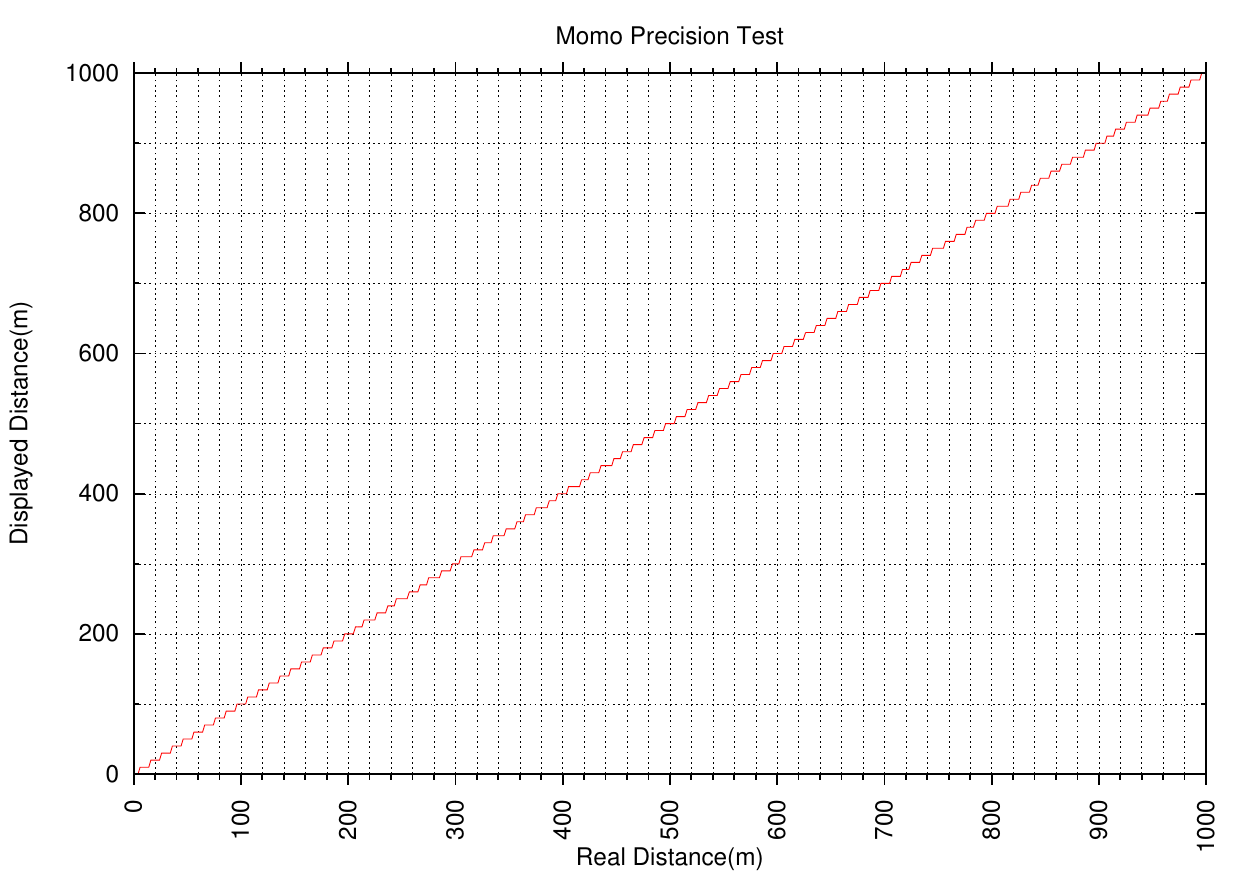}
    \caption{Momo precision test}
    \label{fig:momo-precision}
  \end{subfigure}
  \begin{subfigure}[htbp!]{0.3\textwidth}
    \centering
    \includegraphics[width=\textwidth]{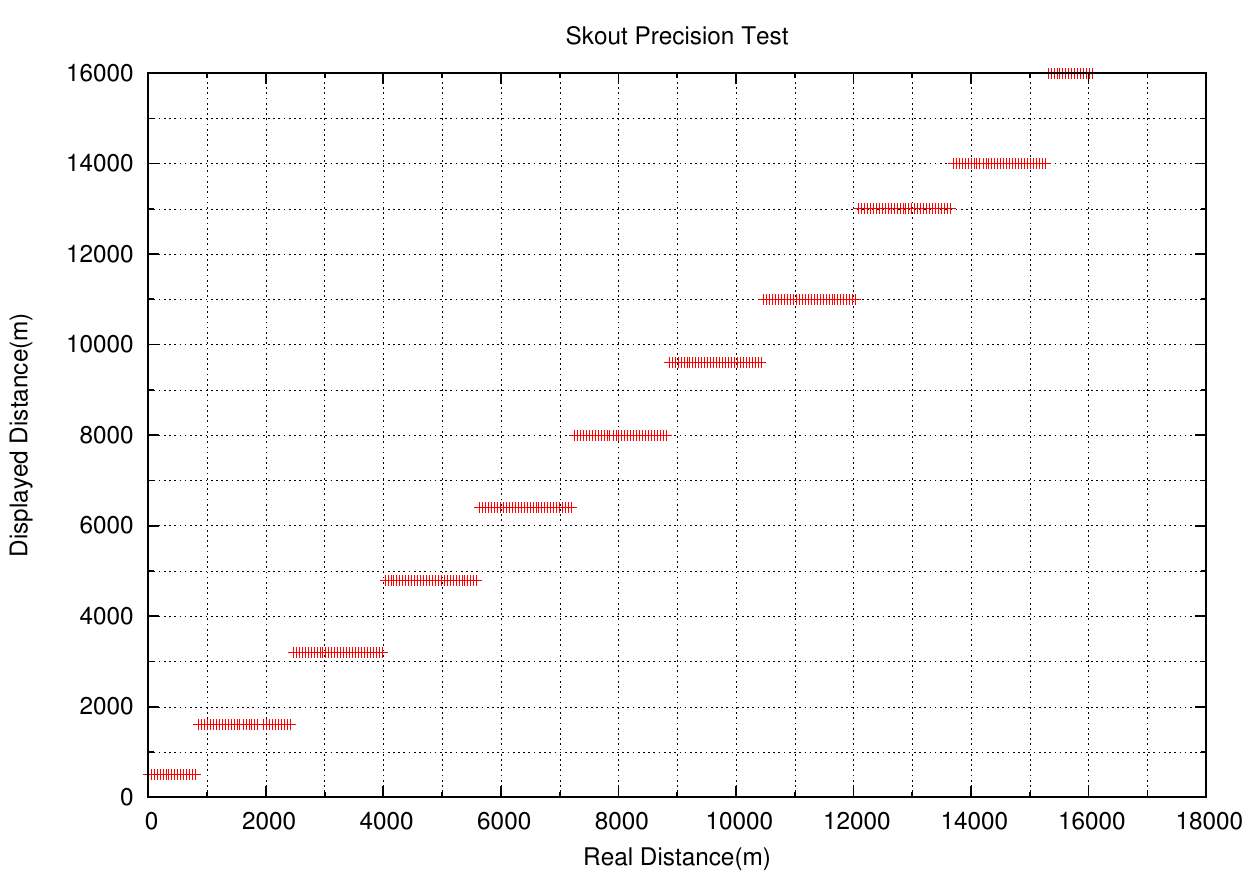}
    \caption{Skout precision test}
    \label{fig:skout-precision}
  \end{subfigure}
  \begin{subfigure}[htpb!]{0.3\textwidth}
    \centering
    \includegraphics[width=\textwidth]{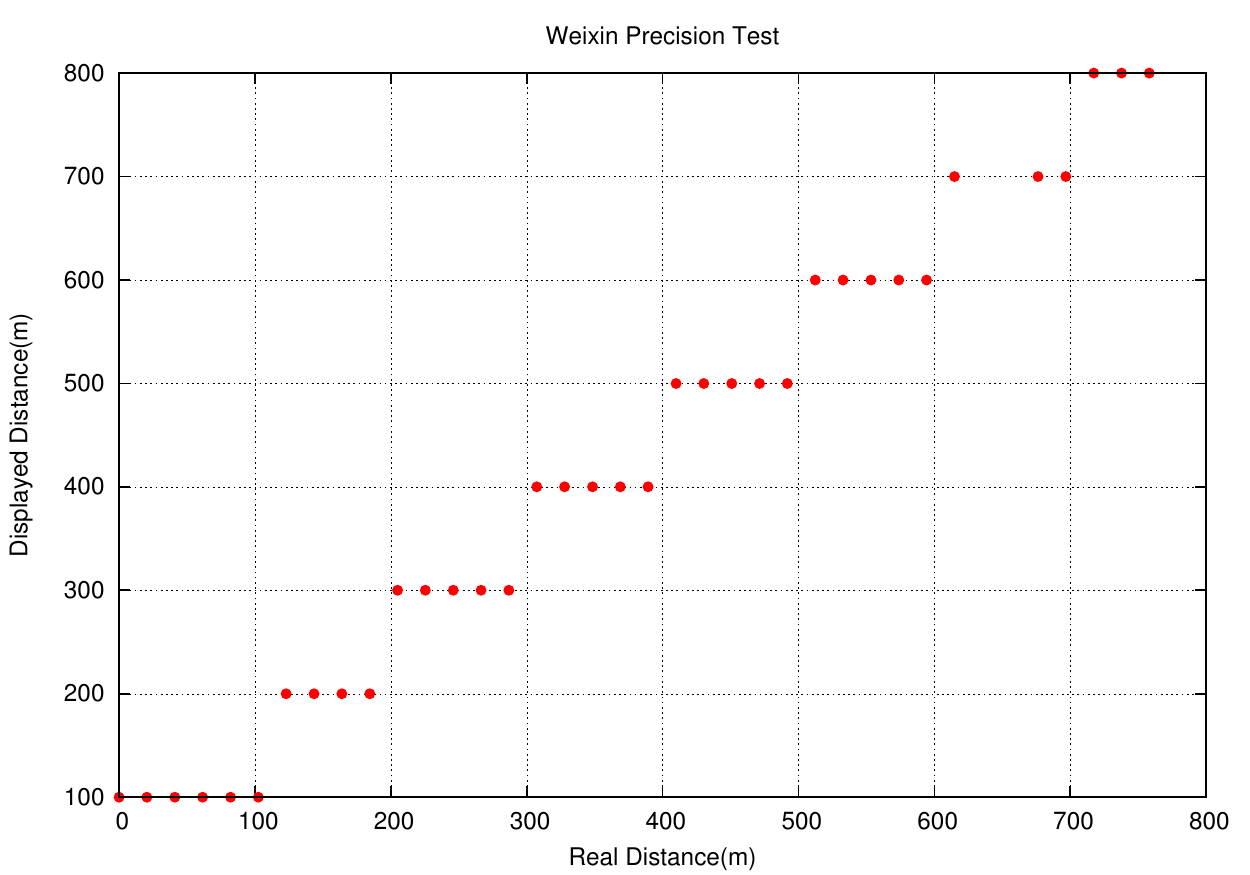}
    \caption{Wechat precision test}
    \label{fig:wechat-precision}
  \end{subfigure}
  \caption{Updating Strategy Evaluation Results}
  \label{fig:app-precision}
\end{figure*}

\subsection{Location Update in LBSNs}
In general, the localization accuracy of smartphone
relies on which kind of location data sources it used. The location data sources (also called
as location providers) include: GPS, WiFi, and Cell ID (cell tower), which could achieve
the localization accuracy of $10$m, $80$m and $600$m, respectively, as shown in the existing works \cite{zandbergen2009accuracy}. However, the location accuracy of location providers will not be
immediately translated into the location accuracy of LBSN apps, which
is caused by different location updating strategies of LBSN
apps. In practice, it's up to the app developers themselves to decide which
location source to trust and it is always a trade-off between waiting
time, precision and energy consumption\cite{android-loc-acc,
  ios-loc-strategy}. To have a better understanding on the updating
strategy of LBSN apps, we perform the following accuracy testing experiments.

In our experiments, we choose GPS localization in which it could
achieve the highest localization accuracy. We mainly perform the
accuracy testing on three apps: Wechat, Skout and Momo. To perform the
experiments, we pre-define a reference point both in the physical
world and the virtual machine. In the apps, this reference point will
be a virtual user located in this position. Then, we enlarge the
physical distance between our mobile device and the reference point
and record the relative distance displayed on apps. We compare the
physical distance and the distance shown in apps and obtain the
accuracy testing results, which are depicted in
Fig \ref{fig:app-precision}.

Since Momo's localization accuracy limit is set to $10$m, we choose a
test point for every $2$m. From Fig \ref{fig:momo-precision}, we
could confirm $10$m as the localization accuracy limit. Such a
distance will be rounded for every $5$m. For example, the distance in
momo will display $0$ if the physical locations of two users are less
than $5$m away. In Skout, the localization accuracy bound is $800$m or
$0.5$mile. In the experiment, we evaluate the localization accuracy
for every $50$m. From Fig \ref{fig:skout-precision}, it is observed
that Skout's minimum coverage is around $800$m, which is $0.5$
mile. Also, the distance will be rounded for half a mile and the
distance will be increased for every $1.6$km or $1$mile. In Wechat,
the coverage bound could be $x$m, where $x$ can be up to $10$km in
sparsely populated area and generally $1000$m in densely populated
places. In the experiment, we set the reference point for every $20$
meters. From Fig \ref{fig:wechat-precision}, it is observed that
Wechat has no round-offs in its distance and the boundary is quite
clear between every $100$m.

\subsection{User Location Privacy in LBSNs}


From above discussions, we could conclude that, in general, the locations
reported in LBSN apps honestly reflect mobile users' real locations, though
users' exact locations are hidden or obfuscated by various location hiding
techniques. For example, Momo only adopts the strategy of showing the
relative distances (strategy I). Skout only shows the distance and, at the same time,
enforces the minimum localization limit (strategy I \& II). As a comparison,
Wechat adopt all of location hiding strategies I, II, III.


In this paper, we argue that, relying on the above mentioned location privacy hiding techniques may
introduce a more dangerous location privacy leaking issues. Due to trust on these location hiding/obfuscating
techniques, LBSN users are more willing to share the PROTECTED
location information with the potential adversary, which could recover
users' exact location or even traces by using the methodology proposed
in this paper. Without full knowledge of its potential risk, LBSN
users may face the serious location privacy leaking issue, while the
adversary could gain a significant advantage during the attack process
since the victim even has no idea about its risks. From the attacker's
point of view, he aims to make an involuntary geo-localization or even
tracking towards a specific victim. In the next section, we will present
our attack methodology in details.



\section{Attack Methodology}
\label{sec:method}

In this section, we will introduce our attacker model, as well as the
attack methodology in details.

\subsection{Attacker Model}
In this study, we consider a capability-restricted attacker aiming
at geo-locating an LBSN user, who does not need to have a priori
social association with him, i.e., an in-app friend. The attacker's
capability is restricted in sense that I) It only has the access right
no more than a normal user of a given LBSN service, which means that
he can only access the publicly available information provided by the
LBSN app. II) It is not allowed to hack the LBSN service by
interfering its internal operations, that is, we do not consider an
attacker that can compromise the LBSN servers and thus can directly
access the user location information as a consequence. In summary, our
attacker is a very weak one which can't gain any additional
information from the LBSN services other than what is entitled to a
regular service user. Specifically, the attacker will try to infer a
user's location information based only on the relative distance
information displayed by the LBSN apps. Note that, to obtain the
relative distance, it is even not necessary for the attacker to be
friend of the victim. Instead, it will automatically display the
relative distance of nearby users in most of considered Category II
apps. For Momo, the attacker can obtain the distance by searching the
victim via Momo ID and in Skout, the distance between the victim user and the
attacker can always be displayed as long as the attacker has sent a regular
message (a greeting for instance) to the victim before.
We are concerned that if the LBSN under examination can't resist even such a
weak attacker, the user's location privacy is obviously in a great
danger as any user can be an attacker.

We further distinguish two different types of attackers, i.e., a
Casual Localization Attacker and a Determined Tracking Attacker. A Casual Localization Attacker reviews
the profiles of nearby users when logging in to a LBSN app as a regular
user, randomly picking up a tracking target and then try to geo-localize the
target.  A Determined Tracking Attacker may start with a known User ID (ID)
and/or User Number as its chosen attacking target and perform the tracking
towards a specific victim for a certain duration. The goal of the tracking attacker
is revealing users' Top N locations (e.g., his home or office)\cite{12-zang2011anonymization}. Note that, a tracking
attacker may start with a target person in mind and exploit certain
side-channel information of the target to help obtain the
corresponding UID or user number. For example, user photos shared
among various social network sites can be used to establish the
linkage for the same user, which in turn can lead to the acquisition
of UID/ User number information in a particular LBSN. Social
engineering approaches like this have been widely studied in the
literature and is not a focus of this paper\cite{iasons2012face}. We assume that a
determined tracking attacker will be able to start with a chosen UID or user
number he wants to locate.

\begin{figure}[htbp]
  \centering
  \includegraphics[width=0.40\textwidth]{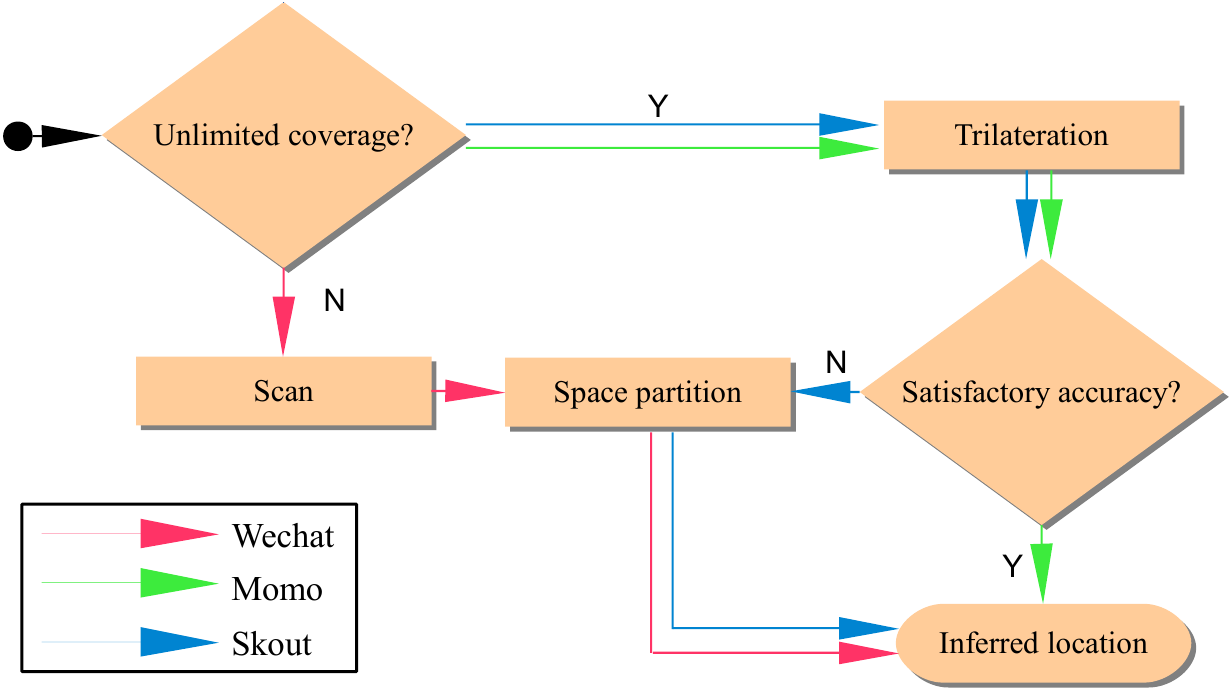}
  \caption{The Attack Flow}\label{intro}
\end{figure}

\subsection{Methodology Overview}
The security of the state-of-the-art privacy protection techniques are based on
the assumption that the location cannot be faked. Under this assumption,
the exact location of the mobile users could be hidden/obfuscated by the above
mentioned three strategies. Therefore, the intuition behind our attack is that, if the attacker could freely generate the fake anchor points with new locations, LBSN apps will be a distance oracle, which will always return the relative
distance with these anchor points to the attacker. By exploiting the
returned information, the attacker could launch different
localization algorithms to geo-locate the victim and even break the
accuracy limit.

As shown in Fig \ref{intro}, the attacking procedure could be illustrated as follows. When the
attacker determines a particular victim, it could generate three fake
anchor locations and obtain the relative distance to the
victim. With $3$ anchor locations and their corresponding distances,
it could trigger the Iterative Trilateration based Localization
Algorithm and obtain the first inferred location, which will be set to
the new anchor point. With this inferred location as well as two other
anchor points, the attacker could launch a new round of attack.  This
process will be repeated until the distance between the new inferred
location and the victim reaches the localization accuracy limit. After
that, the attacker could trigger space partition attack, which further
improves the accuracy until the distance reaches the predefined
accuracy threshold. For those apps with the coverage limit, the
attacker could scan the possible locations until the victim is shown
in the ``nearby list'' of the attacker. Then, it could take advantage
of space partition attack to make an accurate localization. In the
following sub-sections, we will introduce each basic algorithm one by
one.

\subsection{Iterative Trilateration based Localization Algorithm: Skout and Momo}
Our localization approach is based on the traditional Trilateration
Position Problem. In our long distance tracking, we start from 3 randomly generated
positions, which serve as the first three anchor points. In Section \ref{sec:fake-anchor},
we will introduce how to generate the fake locations on Android. The triggered
Trilateration algorithm will return back the first localization results. To minimize
its distance from the target, the least squares solution could be used
to solve this problem as suggested in \cite{liu2006robust}.
We iteratively perform the trilateration localization and generate the next reference point from
the previous round localization results. We denote $P$ as the List of
reference points sorted by the relative distance to the target point
from smaller to larger. Without loss of the generality, the first
three items of $P$ are represented by $p_1, p_2$ and $p_3$. We further
define function $dist(a, b)$ to measure the distance between the point
$a$ and $b$, as well as function $Lsp(a, b, c)$ to return the least
square estimation of the localization target based on three reference
points $(a, b, c)$. We summarize our iterative trilateration
localization algorithm in Algorithm 1. In FreeTrack, the least square
solution is implemented by calling GNU Octave's ``bfgsmin'' method
inside the ``optim'' package. The connection
of Octave and the attacking kit is established by the open source
project Java Octave.



\begin{algorithm}[h]
  \SetAlgoLined \KwData{List $P=\emptyset$, in which the elements are
    sorted by their distance to targeted node}
  \KwResult{$p_0=(x_0,y_0,z_0)$ the location of the target} \BlankLine
  Generate 3 random reference points and put them into $P$\;
  \While{$|dist(p_1,p_3)|>threshold$}{ $p_1,p_2,p_3\leftarrow $ first 3
    elements of $P$\; $t\leftarrow Lsp(p_1, p_2, p_3)$\; Insert $t$
    into $P$\; } Output $p_1$\;
  \caption{The Iterative Trilateration based Localization Algorithm}
  \label{alg:basic-loc}
\end{algorithm}

\begin{figure}[htpb]
  \centering
  \includegraphics[width=0.4\textwidth]{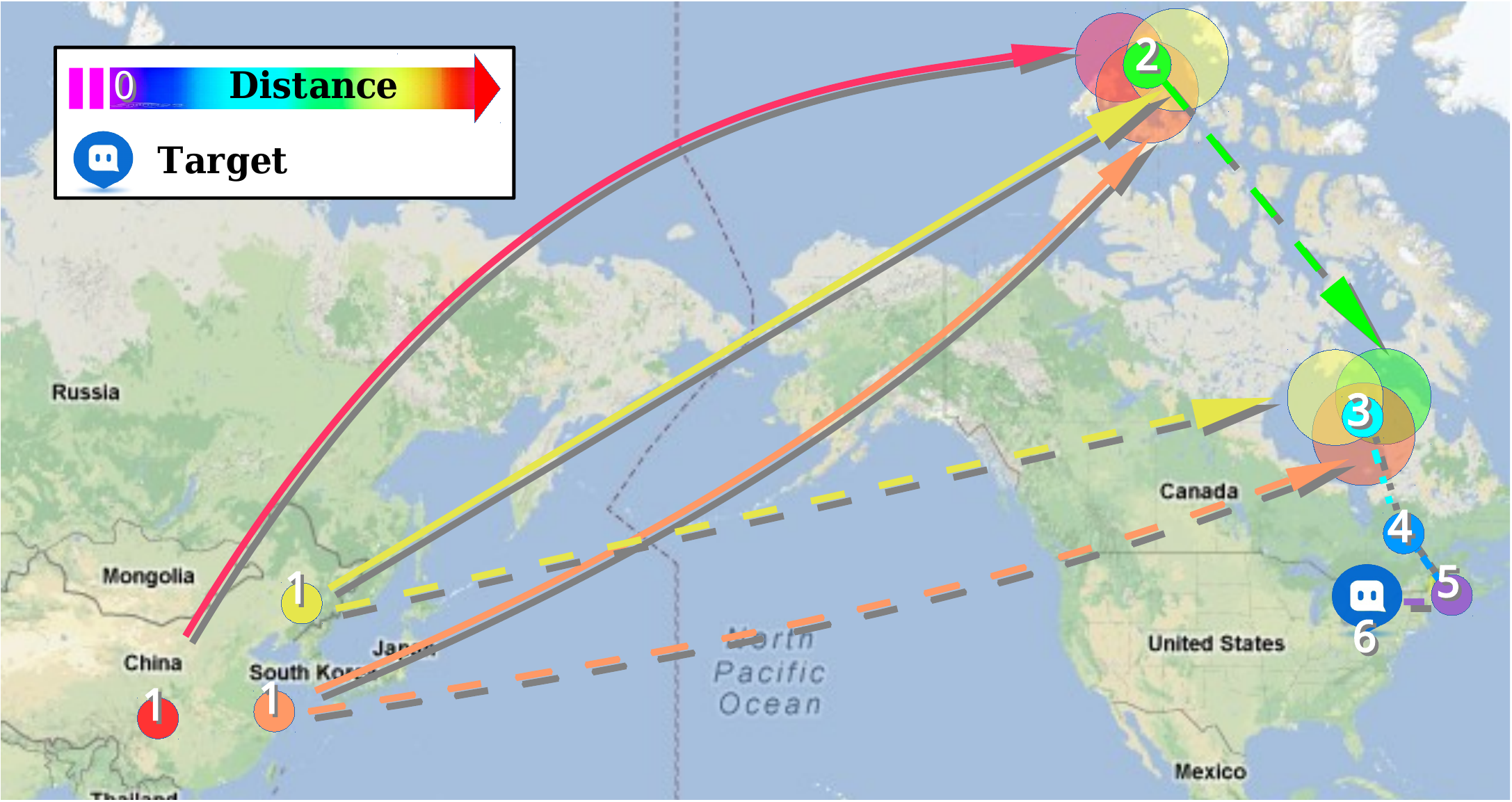}
  \caption{Trilateration on Global Scale}
  \label{fig:global-trilateration}
\end{figure}

\textbf{A Real-world Attack Example}: Due to no display distance boundary, Momo and Skout users
could always obtain their distances with their friends even in the global scale. Fig \ref{fig:global-trilateration}
shows a real-world attack example launched from China towards a user in Bufflo, NY.
Our initial $3$ anchor points are randomly set at Beijing, Shanghai
and Chengdu.  The numbers in the graph represent the positions
inferred in the order. Postions $2$ -- $5$ are intermediate results
and each one is closer to the target. It takes $5$ rounds to finish this attack.

%

\subsection{Breaking Minimum Distance Limit via Space Partition Attack: Skout and Wechat}
\begin{figure}[htbp]
  \centering
  \includegraphics[width=0.23\textwidth]{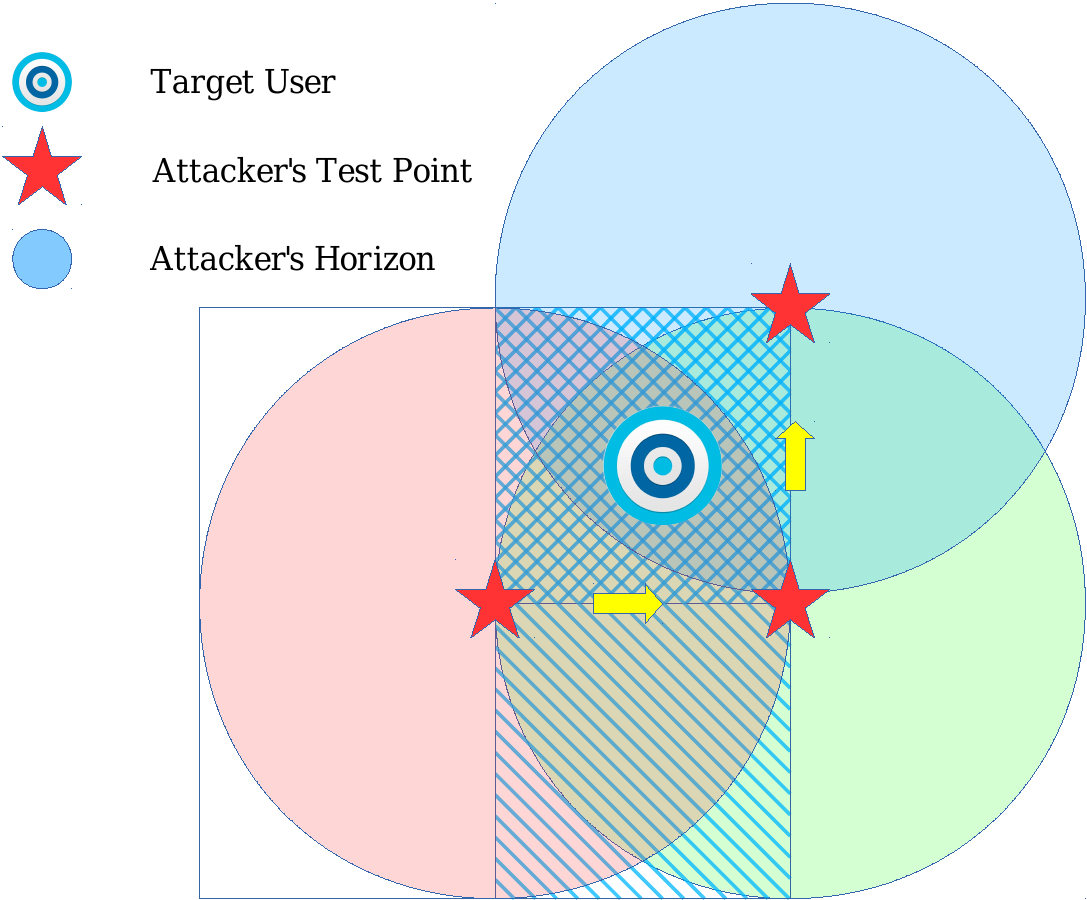}
  \caption{Illustration of Space Partition Attack}
  \label{fig:space-partition}
\end{figure}

Another best practice measure to provide location privacy protection
for users is to limit the relative distance to a certain accuracy,
(e.g., 800 m in Skout or 100 m for Wechat).  In this section, we propose a space
partition attack algorithm to further enhance the localization
accuracy and thus breaking the minimum distance limit. The basic idea of space partition attack is similar to space partition algorithm, which is defined as the process of dividing a space
(usually a Euclidean space) into two or more non-overlapping regions
and thus locating any point in the space to exactly one of
the regions. The basic idea of space partition attack is illustrated
in Fig \ref{fig:space-partition}.

For the simplicity of problem presentation, we consider the minimum
distance limit as the box rather than the circle. Given the minimum
distance limit $R$, the edge length of the box is set to $2R$. The
space partition attack could be illustrated as follows. For each round
checking, the potential area of length $r$ is partitioned into two
regions. Then, we will check if it is within one region. If yes, it is
derived that the user is in this half. Otherwise, the user is located
in another half. We could repeat this partition for multiple rounds
until the expected detecting accuracy is achieved. The whole algorithm
is summarized in Algorithm \ref{alg:partition}.

\begin{algorithm}[h]
  \SetAlgoLined \KwData{An estimated point $p_0=(c_X,c_Y)$
    and its range from target point $T$, given in form
    $dist(p_0, T)\leq R$.}
  \KwResult{$T'$, the final estimation for $T$}
  \BlankLine $dim=X$\;
  $\delta_X=R$\;
  $\delta_Y=R$\;
  \While{$\delta_X \geq threshold$ or $\delta_Y \geq threshold$} {
    Shift $p_0$ in $dim$ dimension by $R$ to $p'$\;
    \If{Distance reading from app $\leq R$} {
      $c_{dim}$=$c_{dim} + \delta_{dim}/2$\;
    }\Else{
      $c_{dim}$=$c_{dim} - \delta_{dim}/2$\;
    }
    $\delta_{dim}=\delta_{dim}/2$\;
    $dim=\{X, Y\} / dim$\;
    $p_0=(c_X,c_Y)$\;
  }

  Output $p_0$\;
  \caption{Space Partition Attack Algorithm}
  \label{alg:partition}
\end{algorithm}

\subsection{Breaking Localization Coverage Bound with Scan and Space
  Partition: Wechat}
Some apps such as Wechat set a certain coverage limit. For such kind
of apps, the first step of launching an attack is to enable the
FreeTrack to see the attacking target shown in the ``Nearby'' list of
this app. To achieve this, if knowing the possible visiting areas of
the target, we could \emph{SCAN} the areas to discover the presence of
the users.  After that, FreeTrack could launch the Space Partition
attack similar to the second stages of other apps.


The scan strategy is to query over a particular area at a particular
distance $d$ until the user is presented inside the ``Nearby''
list. Take Wechat as an example, in a high user density area (e.g.,
Lujiazui Area of Shanghai), to cover an area of $28$km$^2$ with the
distance $d=1$km, it only needs to query $28$ times at the worst case.
It is noted that, the coverage limit of Wechat varies with the user
density of this region.  In an area of low user density (e.g,
Buffalo), the coverage bound could be as far as $10$km.  Therefore,
it only takes $5$ queries to cover the whole downtown area. Upon
discovering a user, we could use Space Partition attack as indicated
in last section.

\textbf{Performance Enhancement by Using Social Popularity Index:}
To further improve the performance of the launched attack, 
we could use the social popularity index to accelerate the attack. 
This strategy comes from a simple observation that, the possibility of
a node staying at different locations is far from uniform, and it is very likely that a user
stays at the location that is most popular at that time. The social
popularity is a spatial-temporal concept. Intuitively, it is more
likely for a user to be at the restaurant at 6 pm in the afternoon
rather than at office. However, this statement may not hold at 10 am
in the morning. In our implementations, we measure the social
popularity of different locations by collecting their user population
information at different time slots. Then, based on the number of
users, we could assign a higher priority to those areas with the
higher user population.

In addition to the above described methodology, there are still
implementation challenges, including: how to generate the fake
locations in smartphone which allows the above methodology to work and how to fetch the relative distance after setting a fake
location in smartphones. More importantly, all of these should be performed
in an automated way. We'll introduce the implementation details in the next section.

%

\section{Implementation of FreeTrack}
\label{sec:implementation}
Besides the algorithms introduced in previous section, the
implementation of FreeTrack will require other 2 key modules: the
location spoofing module and the location reading module.  Mainly, our
FreeTrack is implemented in Clojure\cite{clojure-web} in order to cope
with MonkeyRunner\cite{monkey-runner}
to control Android virtual machines and send commands. We also
implement a LocationFaker app that receives HTTP request from
FreeTrack and set the location in Android. To address problems we
encounter during location faking and result reading, we make multiple
tweaks in the Android framework as well.

\subsection{Generating Fake Anchor Locations on Android}\label{sec:fake-anchor}
To launch the proposed attack, we need to allow the FreeTrack to
freely generate the anchor location points, which are used to obtain
the relative distances of the victim users. Android SDK ships with
QEMU based virtual machine that allows setting location via Telnet,
but the virtual machine is too slow for real-world automated tracking. Thus, we set our Android
system on real Android X86 images\cite{androvm} running on
VirtualBox\cite{virtualbox}. It is important to point out
that, since almost all of the LBSN apps cover all of the platforms
(please refer to Table \ref{tbl:apps} for our survey),
including IOS/Android/WP, spoofing the android device's locations
could allow the attacker to obtain the relative distances with LBSN
users on IOS/WP, and thus launch the attack towards the users on all
of platforms.

There are several ways to spoof Android device's locations, including:
using mock location, or intercepting network traffic. We do not consider
some existing apps such as Developer Shell \cite{developer-shell} and FakeGPS \cite{fake-gps} because they are
either unstable due to potential bugs or cannot allow us to set the
location arbitrarily, which motivates us to implement our location
spoofing component, LocationFaker. LocationFaker is implemented as
a system service which eliminates the possibility that Android system
may kill the activities to release resources. Also, it has embedded Jetty\cite{jetty} as a web server to
provide stateless http-based interface to set locations and act as a
fake location server when we redirect the network traffic.


In general, most of the LBSN apps on Android either use built-in
Android API (Wechat or Skout) or third-party SDKs (e.g., Momo using
Baidu Location SDK), which lead to different spoofing strategies.
For
built-in Android API, we adopt \emph{Fake Location Provider based
location spoofing}. However, according to the official
document\cite{baidu-sdk-faq}, Baidu Location SDK does not function
well on virtual machines. For this case, we achieve location spoofing
by using \emph{Network Redirection}. In the follows, we will introduce
both of approaches in details.

\subsubsection{Location Spoofing with Fake Location Provider}
Android apps mostly acquire locations via one or more location
providers (e.g., ``gps'' and ``network'') retrieved from the \emph{location} system service. Android allows users to
freely add location providers under certain circumstances such as
debugging or providing locations from other devices, eg. Bluetooth.
By enabling ``Allow mock location'' option in developer options and
adopting the API ``addTestProvider'', it is possible to add a
user-written location provider. Interestingly, we could set the
provider's name to ``gps'' to make it indistinguishable from the real
gps, and thus fool the system and make it believe that they are
receiving locations from the real GPS chip.  Our fake location
provider is running on its thread, feeding location information every
$700ms$. 

Another challenge of spoofing the location on Android is that the
provided location should satisfy a certain accuracy. If failing to
achieve, some apps may fail to accept it. For example, in Wechat, it
is observed that the system will return error messages if FreeTrack
tries to send the locations to Developer Shell. We verified it by
checking it manually on Google Map and it is found that the accuracy
is only 90,000m. To address this issue, we decompiled the
Android system framework with ApkTool\cite{apktool} and modified the
constructor of ``android.location.Location'' by coercing
``mHasAccuracy'' to ``true'' and enforcing ``getAccuracy'' to always
return $70$m.  In this way, apps will always retrieve consistently
accurate value under different circumstances and the location faking
component starts to work as expected.

\subsubsection{Location Spoofing with Network Redirection}
For those LBSN apps which do not adopt Android built-in APIs for
location retrieval, we introduce another approach based on
Network Redirection. In this section, we use Momo as an example to
show how it works. Basically, Momo uses Baidu Location SDK to obtain
the user location. 
We start from analyzing the network traffic with
Wireshark\cite{wireshark} and Tcpdump\cite{tcpdump}. It is observed that the API first posts the coordinates and supplemental
information, which is obtained from the device, to
\url{http://loc.map.baidu.com/sdk.php}. The server returns a plaintext
JSON object carrying location information as follows:

{\small
\begin{verbatim}
{"content":{"addr":{"detail":""},
            "bldg":"","floor":"",
            "point":{"y":"","x":""},
            "radius":""},
 "result":{"error":,"time":""}}
\end{verbatim}}

By comparing the failed request against successful one,
it is found that the key fields are the $x$ and
$y$ coordinates in ``point'', ``radius'', the error code and the
timestamp. The error code 161 indicates a \emph{successful} query and
the $y$ and $x$ carry the computed coordinates of the latitude and
longitude. We utilize Iptables\cite{iptables} in our implementation to
build a NAT that redirects all the requests originally sent to Baidu
location server back to our embedded Jetty web server
running by LocationFaker. LocationFaker will then construct a similar
JSON object carrying fake locations to trick Baidu Location
SDK to accept the received location as the real location.


\subsection{Fetching the Location Readings}

Location fetching module is the last component of FreeTrack. The basic strategy of FreeTrack is actually running the client and simulating the user's inputs to retrieve distance
readings. To simulate user inputs, we adopt the MonkeyRunner library bundled
with the Android SDK. With MonkeyRunner scripts provided in Jython, it
simulates user inputs in apps to allow us automatically to
perform various tests on apps. We integrate the API with our attacking framework to
allow user defined inputs. We simulate consecutive operations in forms
of touch, drag, scroll, input numbers, shell command and key press to mimic
a user's behavior to the apps to trigger a location information update
and scroll down the list to read out all items.



To read the distance from the apps, we choose to modify the framework
to dump the text in stead of choosing OCR since the former one is more
reliable and accurate especially in virtual machines. To dump the
text, we modified the ``android.widget.TextView'' to dump text to
log messages whenever ``setText'' method call is made.
FreeTrack then retrieves text from the Adb logcat buffer and reads
specific app's output by filtering log level, grepping by PID and tags
then matching particular regular expression pattern.


\section{Real-world Attack Evaluations}
\label{sec:evaluation}
To evaluate the effectiveness of FreeTrack, we implement the
real-world experiments by recruiting $30$ volunteers for the 3 kinds
of LBSN apps: Wechat, Skout and Momo. We evaluate the
\emph{Localization Accuracy} of FreeTrack by comparing the distance
between the user's \emph{Real Locations} and \emph{Inferred
  Locations}, and \emph{Localization Efficiency} of FreeTrack by
measuring the latency of launching an attack for different apps.  In
the experiments of real-world tracking, we evaluate the effectiveness
of FreeTrack by measuring how many top locations could be recovered by
using 3-week track.

\subsection{Localization Accuracy and Efficiency}
To well evaluate the localization accuracy, we set that the attack is
triggered as soon as the user reports his real location obtained from
location providers (e.g., GPS, wifi, or cell ID). The attack and real
location reporting is set to the synchronous mode because we need to
make sure that users' mobility will not impact the localization
accuracy. To achieve this, we deploy a web server in which users with
HTML5-capable browsers could retrieve their locations directly from
location providers of their smart phones, and then submit their real
location, user information to the server. The server will immediately
launch an attack toward this user by using his user information. The
server will maintain a task queue and each idle node will be assigned
with a task and schedule an attack, the results of which will be
reported to the server and compared with the exact location. Members
of our groups regularly submit their locations to the server. We've
collected in total of more than $350$ location reports and attack
results. The testing regions include United States, China and Japan.

\subsubsection{Localization Accuracy}
\begin{figure*}[htbp]
  \centering
  \begin{subfigure}[htbp!]{0.32\textwidth}
    \centering
    \includegraphics[width=\textwidth]{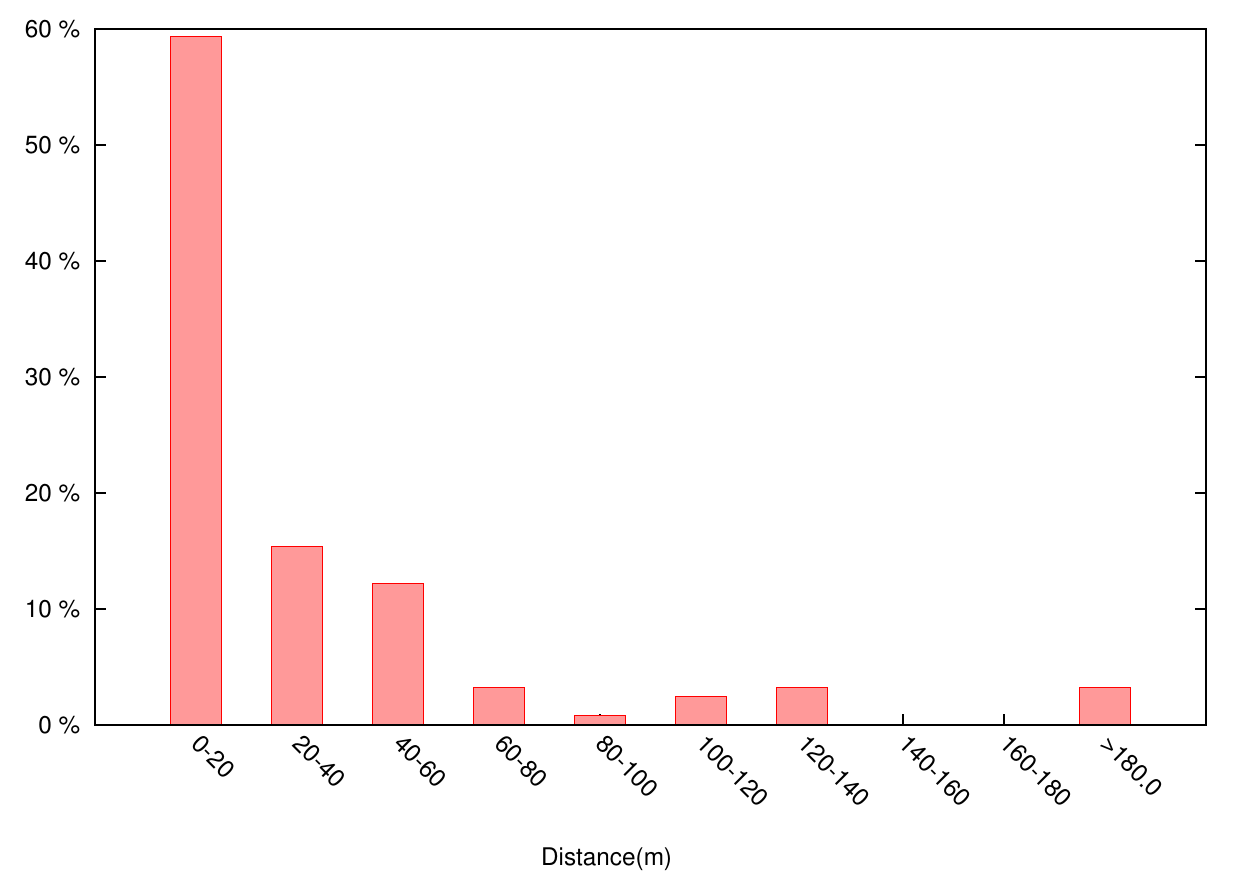}
    \caption{Momo's Localization Accuracy}
    \label{fig:momo-acc}
  \end{subfigure}
  \begin{subfigure}[htbp!]{0.32\textwidth}
    \centering
    \includegraphics[width=\textwidth]{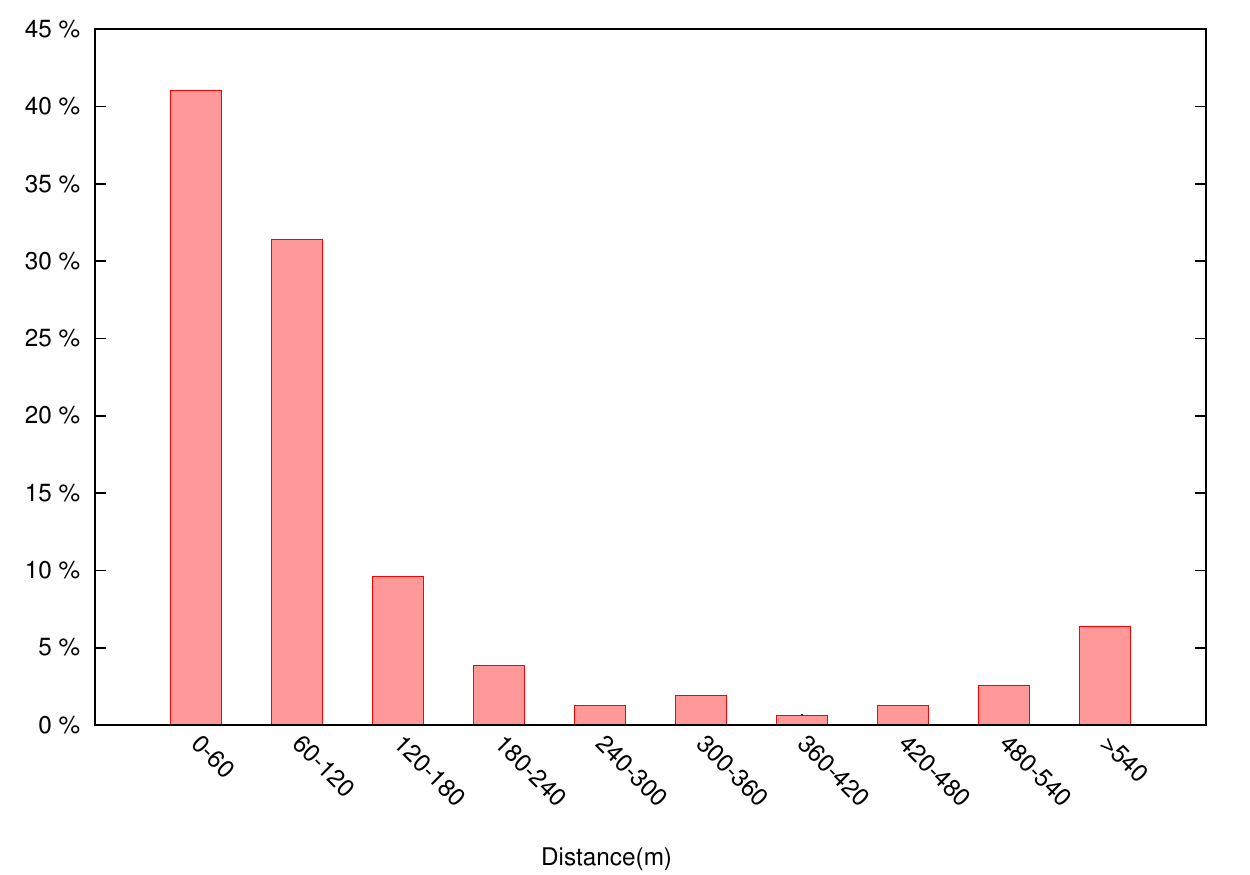}
    \caption{Skout's Localization Accuracy}
    \label{fig:skout-acc}
  \end{subfigure}
  \begin{subfigure}[htpb!]{0.32\textwidth}
    \centering
    \includegraphics[width=\textwidth]{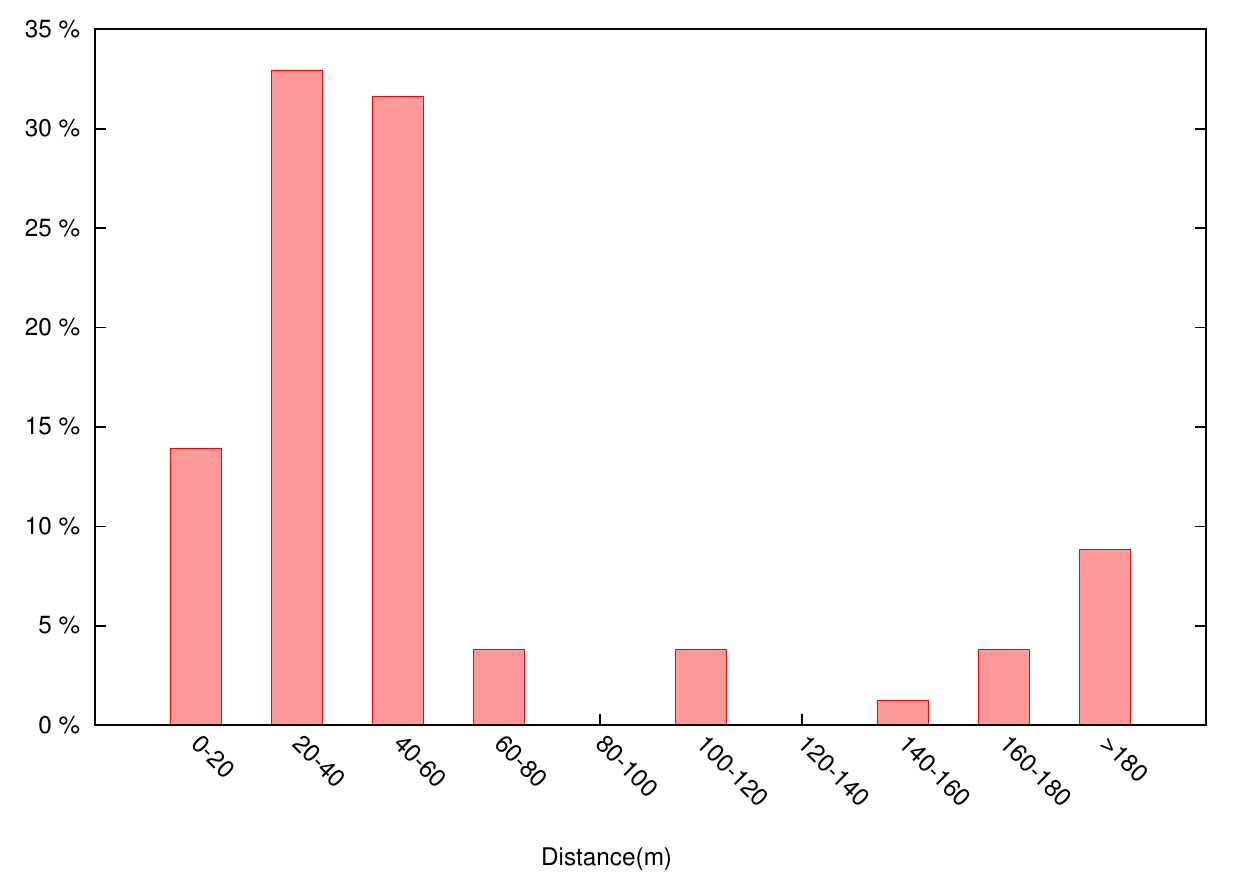}
    \caption{Wechat's Localization Accuracy}
    \label{fig:wechat-acc}
  \end{subfigure}
  \caption{Evaluation on Localization Accuracy}
  \label{fig:app-acc}
\end{figure*}

The evaluation on localization accuracy is shown in Fig
\ref{fig:app-acc}. From Fig \ref{fig:app-acc}, it is observed that the
majority of the results achieve a very high localization accuracy. For
Momo, nearly $60\%$ of the attacks can geo-locate a user at the
accuracy of less than $20$m and only less than $10\%$ of the
localization accuracy is more than $60$m. In general, it could achieve
an average localization accuracy of $25.8233$m for $119$
evaluations. For Skout, though the minimum localization limit is
$800$m, most of the localization could achieve the accuracy of less
than $60$m while over $70\%$ of the localization is less than
$120$m. The average localization accuracy could reach $129.3674$m for
$156$ tests, which well demonstrates the effectiveness of the Space
Partition algorithm. For Wechat, whose minimum localization limit is
$100$m, FreeTrack is able to geo-locate $50\%$ of users in less than
$40$m.  The average accuracy is $51.0888$m for $74$ tests. Note that,
there are different factors which contribute to the localization
errors. For example, localization errors may come from
different way to fetch location from HTML geo API, choosing
inconsistent location providers, various location calculation
algorithm or location cache.

\subsubsection{Localization Speed}
\begin{figure}[htpb!]
  \begin{subfigure}[htbp!]{0.23\textwidth}
    \centering
    \includegraphics[width=\textwidth]{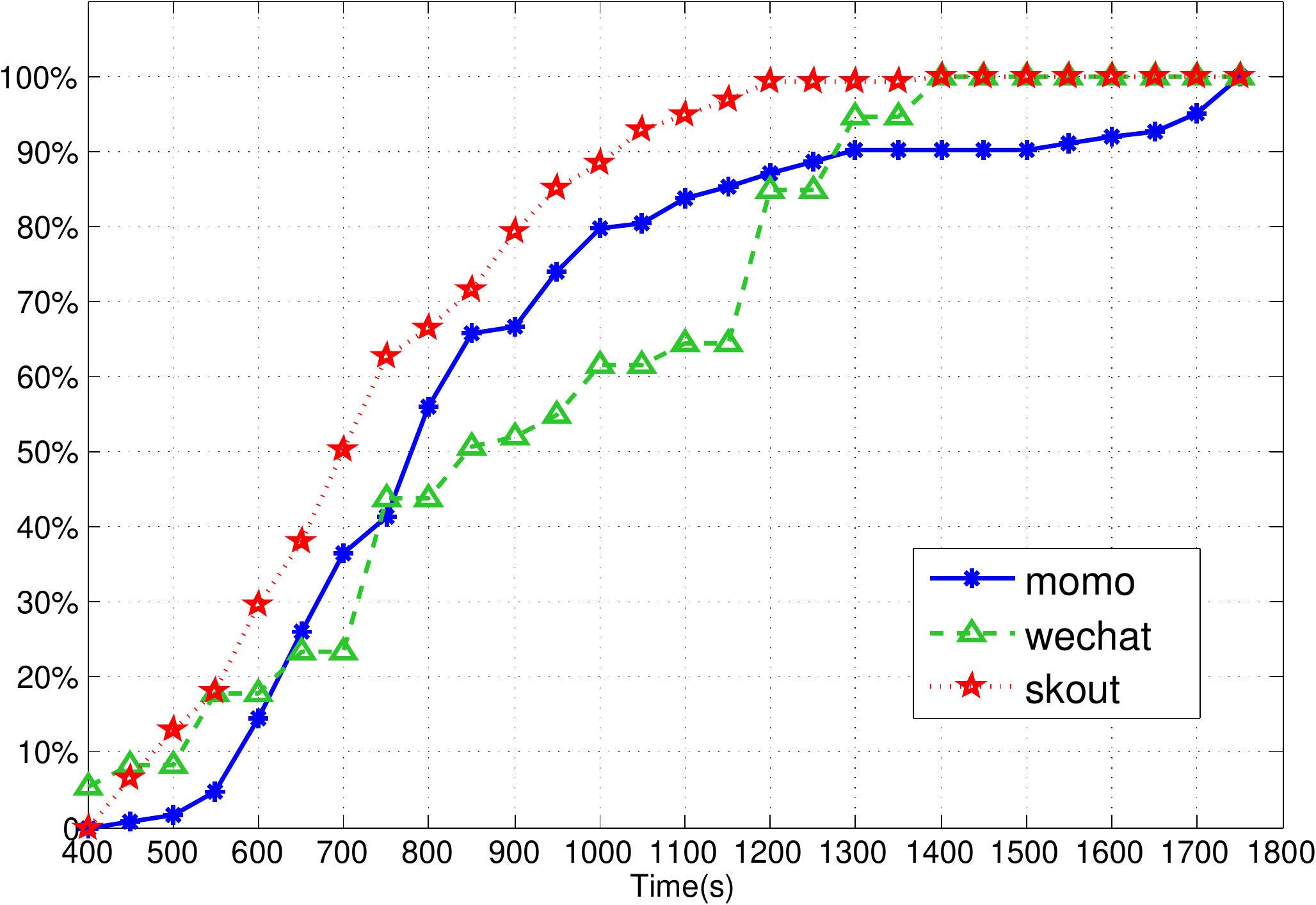}
    \caption{Localization inference time on different apps}
    \label{fig:loc-real-eff}
  \end{subfigure}
  \begin{subfigure}[htbp!]{0.23\textwidth}
    \centering
    \includegraphics[width=\textwidth]{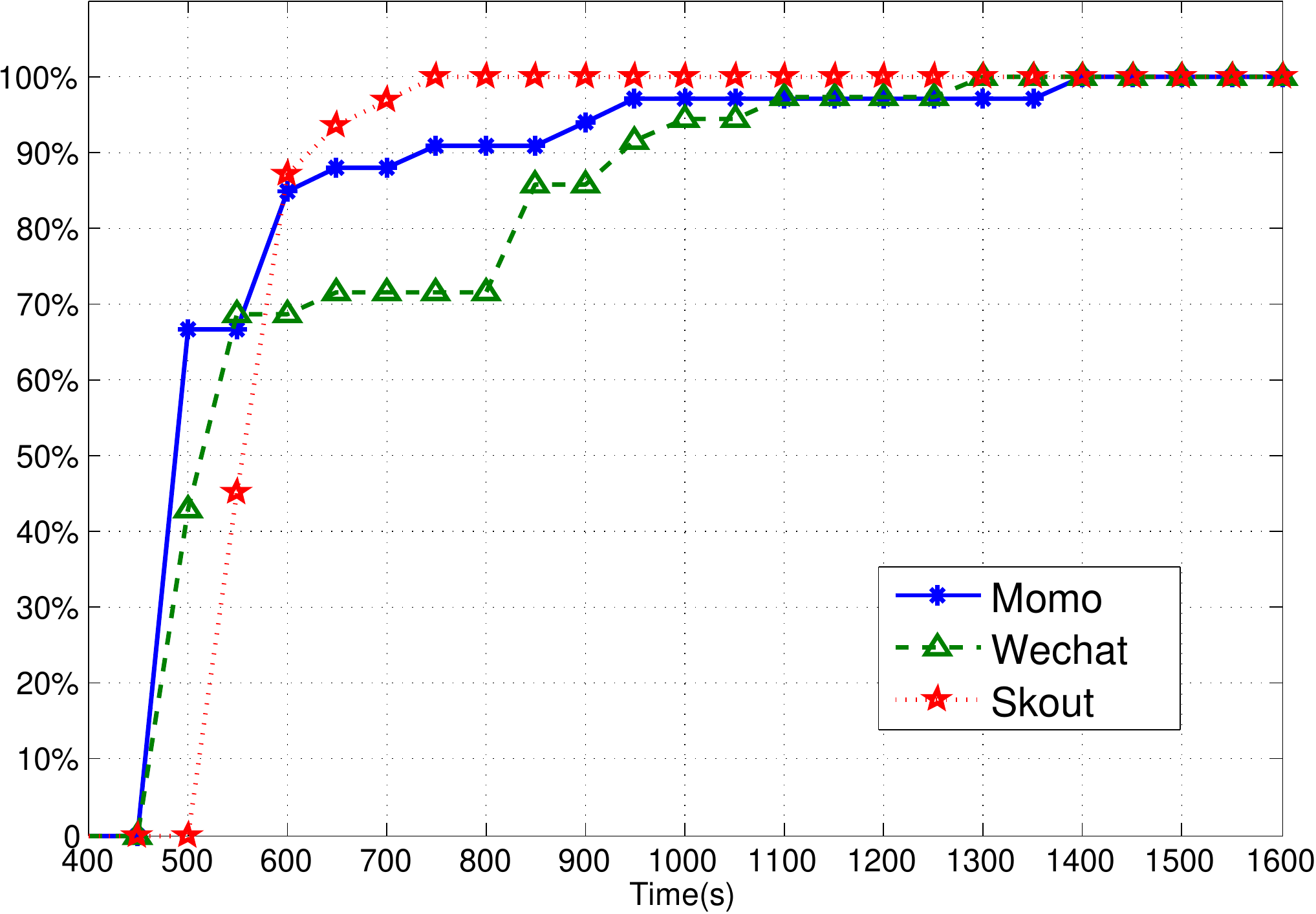}
    \caption{Improved localization inference time on different apps}
    \label{fig:loc-real-eff-imp}
  \end{subfigure}
  \caption{Localization Efficiency of the Original and Enhanced
    Scheme}
  \label{fig:run-time-comp}
\end{figure}

We also evaluate the efficiency of FreeTrack by measuring the
execution time of an individual attack and the results are shown in
Fig \ref{fig:loc-real-eff} and Fig \ref{fig:loc-real-eff-imp}, which
correspond to the case of randomly setting first $3$ anchor points and
social popularity enhanced attacking approach.

From the Fig \ref{fig:loc-real-eff}, it is shown that over $80\%$ of
the attacks for all $3$ apps could be finished within $1200s$. It is
important to point out that most of the time is spent on waiting for
the app server's response. Take Wechat as an example. Each query
should wait $40s$ to ensure that the user's location is fetched due to
network latency and for Momo, the number increases to $55s$ while for
Skout, it spends on $20s$ on queuing per query. In the evaluation,
Momo has a faster localization speed as the iterative trilateration
converges faster than space partition and thus requires less query
time.  From Fig \ref{fig:loc-real-eff-imp}, it is shown that, after
adding some side information such as social popularity index in Wechat
or setting the initialization point in the approximate area (e.g.,
Shanghai) for Momo or Skout, the localization performance could be
enhanced for $1.5$ times.

\subsection{Real-world Tracking: Tracking Accuracy and Top Location Coverage}
\begin{figure}[htpb]
  \centering
  \includegraphics[width=0.38\textwidth]{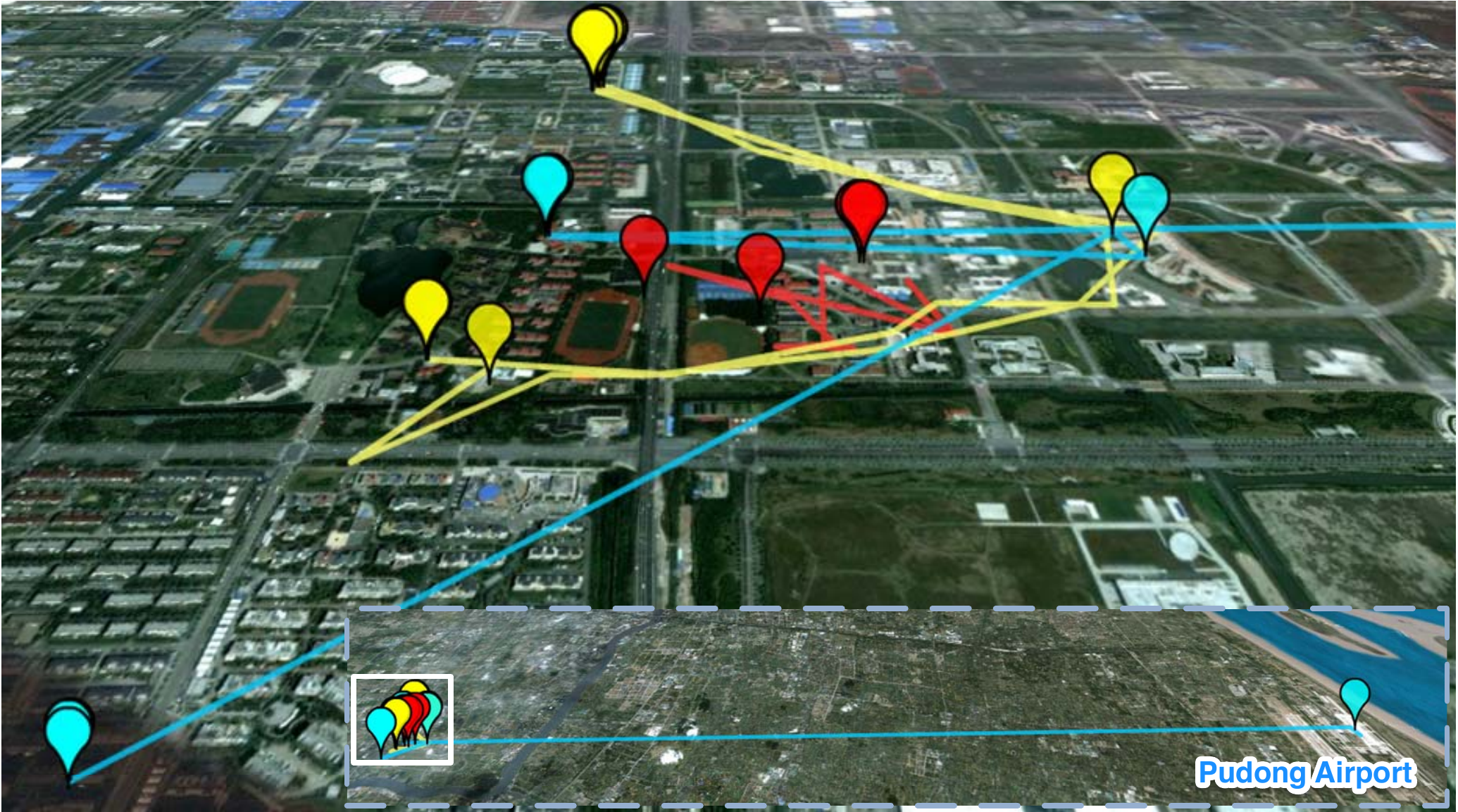}
  \caption{Three Real-world Traces and Inferred Locations}
  \label{fig:real-trace-infer}
\end{figure}

In this section, we evaluate the effectiveness of FreeTrack in
real-world tracking. The basic goal of this experiment is to compare
the inferred mobility traces of the mobile users with their real
mobility trace to measure how much location information the attacker
could obtain by tracking the users in a certain duration. In this
phase, we recruit $30$ volunteers from China, Japan and United States
to participate in our three-week real-world experiments. Due to no
display coverage limit, the tracked Skout and Momo volunteers are
scattered in these three countries. For Wechat, due to the coverage
limit, FreeTrack covers a region of the size of 3km$*$5km in Shanghai
and $20$km$*$20km in Buffalo.  In these
three weeks, the volunteers use the LBSN apps in the same way as other
typical LBSN users. To obtain the ground truth data (or user's real
mobility traces), we develop an app based on Baidu Location API, which
runs as a service in the background, recording their locations every
half an hour and submits the traces to the server. In the server side,
we run $3$ Momo FreeTrack instances, $7$ Wechat nodes and $3$ Skout
nodes to track Momo, Wechat and Skout users, respectively. We
continuously track them for $3$ weeks and collect $3395$ inferred
points in total. Fig \ref{fig:real-trace-infer} shows the tracking
results of 3 users in one day. Note that, the plotted red, yellow, and
blue lines represent users' real mobility traces, while the red,
yellow, and blue bubbles indicate the locations inferred by FreeTrack.

\subsubsection{Tracking Accuracy}
In real-world tracking, synchronization of user real trace reporting
and our tracking is almost infeasible due to unexpected user usage
pattern as well as the randomness of the delay between victim's
location updating and our tracking.
Therefore, we also evaluate the
tracking accuracy in the asynchronous mode. In particular, the user's
real-world trace is periodically updated (e.g., 30 mins), and the
tracking on users is also periodically launched (40 mins). In this
case, we define the \emph{Tracking Accuracy} as the distance of the
inferred location and its closest counterpart of the reported user
traces (ground truth data) in time domain. In general, tracking
accuracy provides the upper bound of the localization error.

\begin{figure*}[htbp]
  \centering
  \begin{subfigure}[htbp!]{0.32\textwidth}
    \centering
    \includegraphics[width=\textwidth]{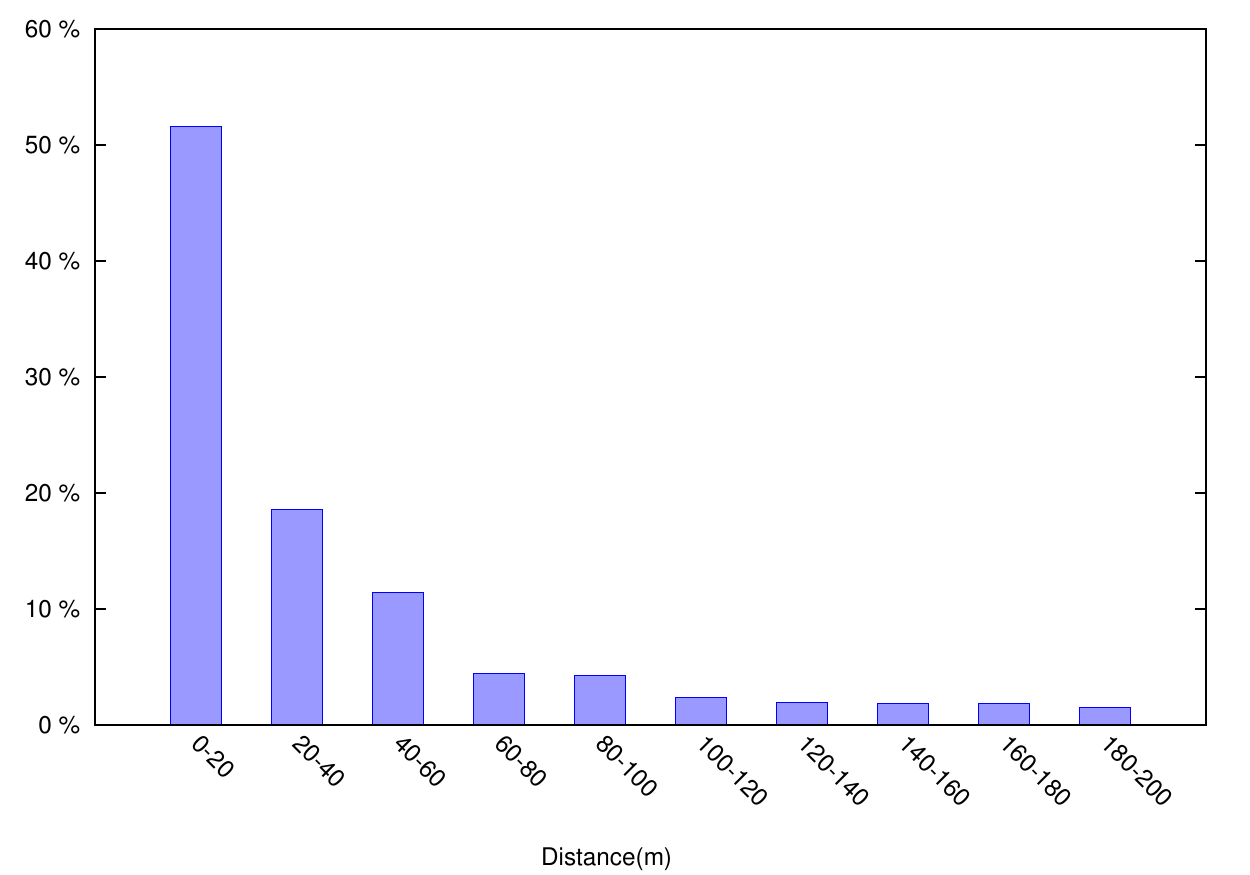}
    \caption{Tracking Accuracy of Momo}
    \label{fig:momo-async-acc}
  \end{subfigure}
  \begin{subfigure}[htbp!]{0.32\textwidth}
    \centering
    \includegraphics[width=\textwidth]{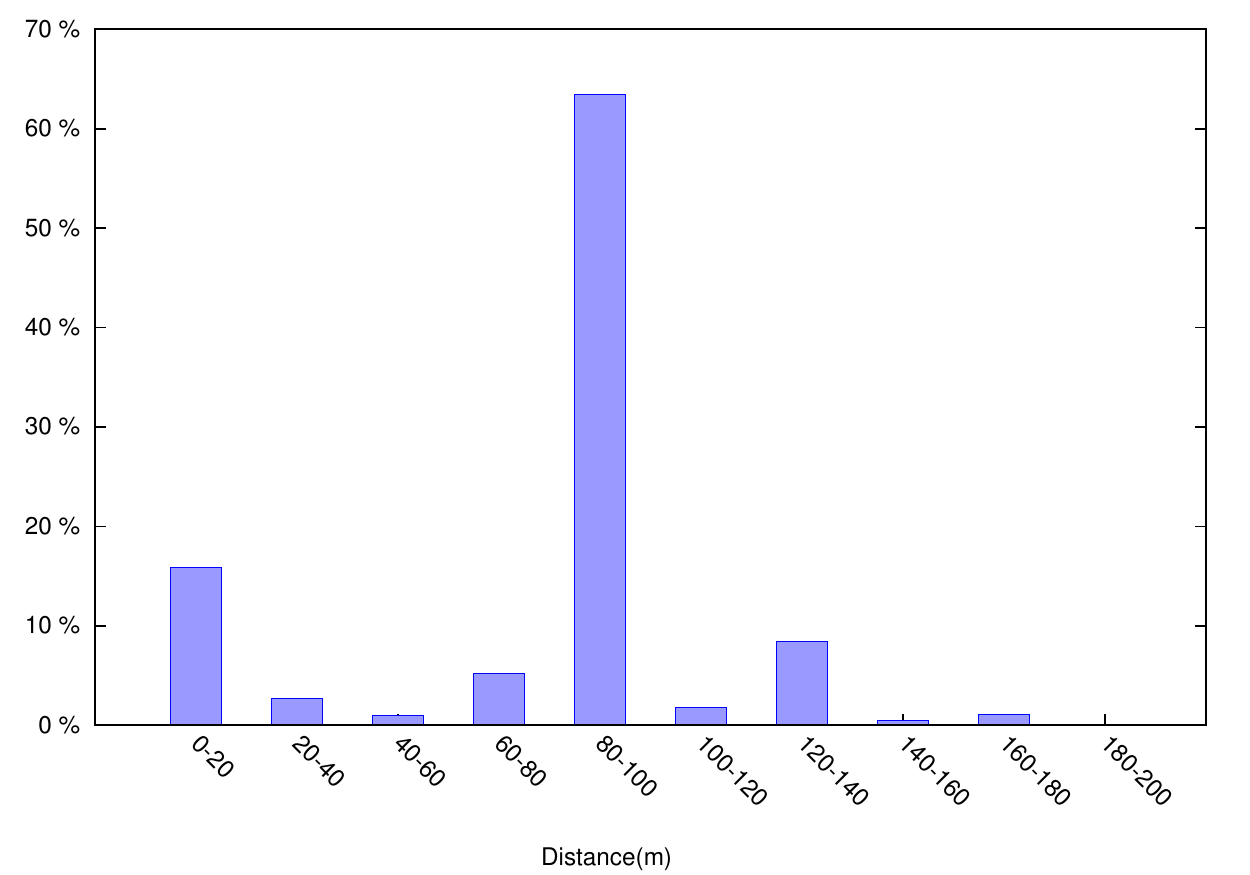}
    \caption{Tracking Accuracy of Skout}
    \label{fig:skout-async-acc}
  \end{subfigure}
  \begin{subfigure}[htpb!]{0.32\textwidth}
    \centering
    \includegraphics[width=\textwidth]{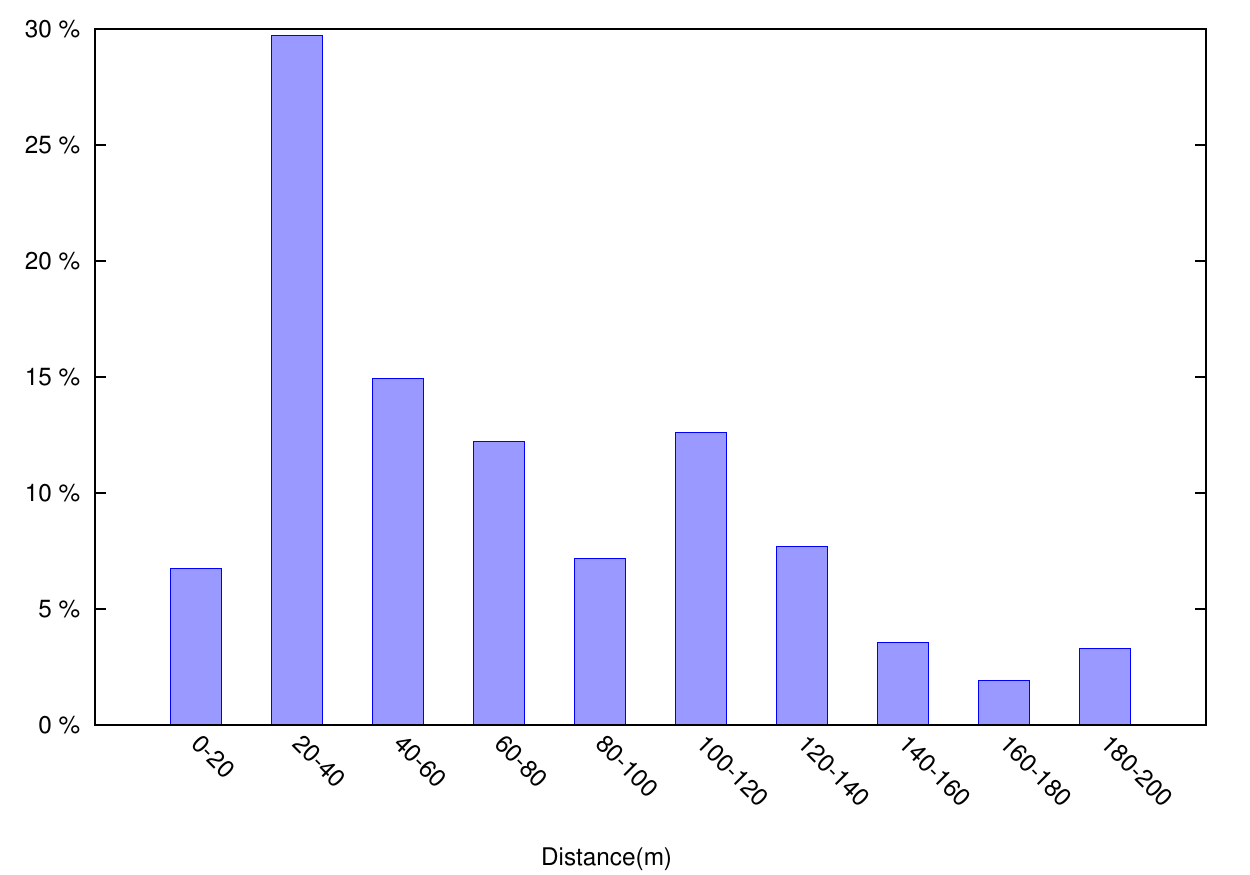}
    \caption{Tracking Accuracy of WeChat}
    \label{fig:wechat-async-acc}
  \end{subfigure}
  \caption{Evaluation Results on Tracking Accuracy}
  \label{fig:async-acc}
\end{figure*}

The evaluation of tracking accuracy is shown in Fig
\ref{fig:async-acc}. The experiment results demonstrate that the
asynchronous tracking can also achieve a very high level of
accuracy. As shown in Fig \ref{fig:momo-async-acc}, more than $80\%$
of tracking results on Momo can geo-locate the victims in 40m, more
than $90\%$ of tracking results on Skout can break the distance limit
of $800m$ to geo-locate the victims to $0-20$m and $80-100$m, and over half
of the tracking on Wechat users can be located to the accuracy of less
than $60$m.

\begin{figure}[htpb]
  \centering
  \includegraphics[width=0.32\textwidth]{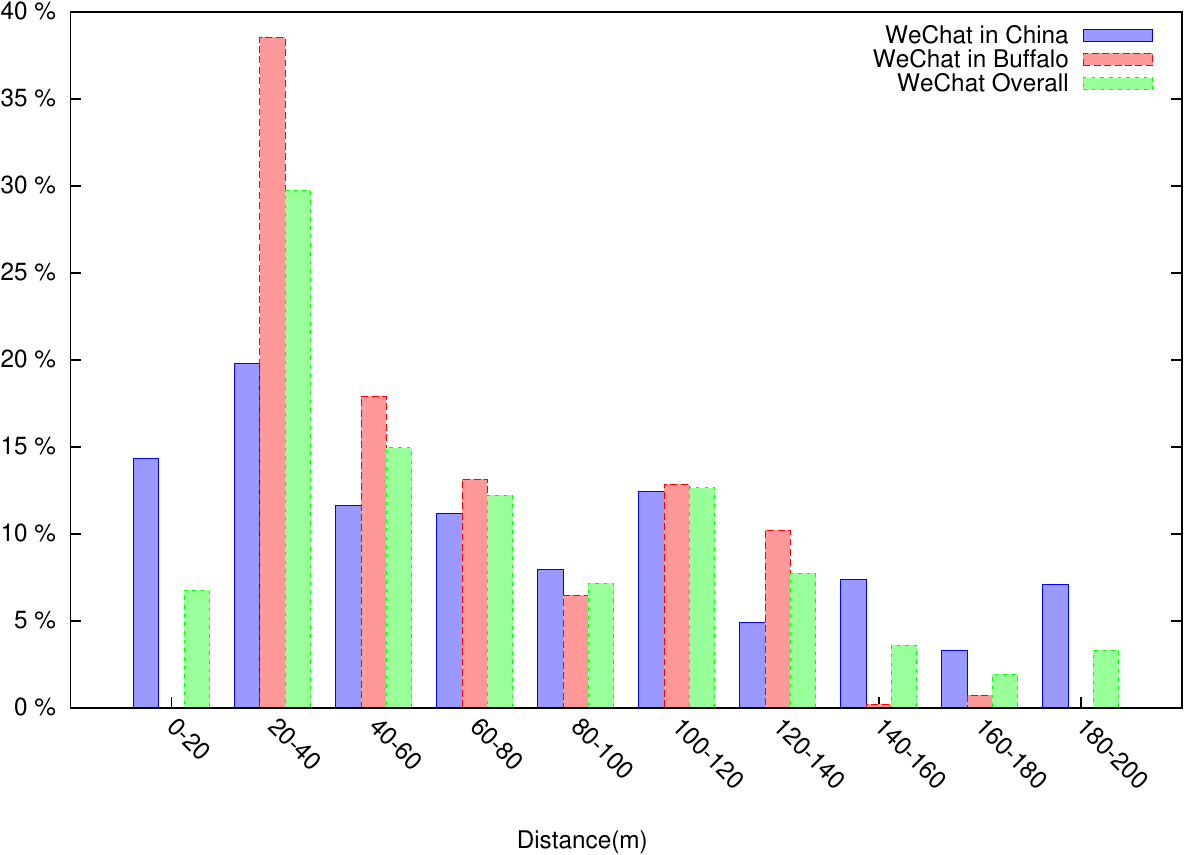}
  \caption{Wechat Accuracy Comparison }
  \label{fig:wechat-acc-comp}
\end{figure}

The factors which may potentially affect the tracking accuracy
includes: the location providers (GPS, Wifi, or Cell ID), which have
different localization accuracy of less than $10$m, tens of
meters, and several hundred meters, respectively; cache policy, which
defines how long the user's location is buffered at the server
side. We've investigated the cache policy of Wechat by comparing the
results from China and US, which have different user populations and
thus different cache time.  The results are indicated in Fig
\ref{fig:wechat-acc-comp}. In China, as one of the most popular LBSN
apps, Wechat has a huge population of users, which makes the users'
locations buffered at the server side for a shorter duration, making
user tracking more difficult to perform in China. Instead, it is much
easier to track a Wechat user in United States due to less number of
users and a much longer time of users' location cache.

\subsection{The Coverage Rate of Top N Location}
According to \cite{12-zang2011anonymization}, ``Top N'' locations
refer to the locations that are most correlated to users'
identities. For example, ``top 2'' locations likely correspond to home
and work locations, the ``top 3'' to home, work, and shopping/school
locations. In the section, we investigate how much location
information the attacker could gain from launching FreeTrack by
introducing the concept of Top N location Coverage Rate, which is
defined as follows. Given $\mathbb{G}$ as the set of reported traces
(ground truth data) and $\mathbb{I}$ as the set of inferred traces, we
define $Top_N()$ as the function that returns $N$ most visited
locations from a specific trace and thus define Top $N$ Location
Coverage rate as
\begin{displaymath}
  TNR=\frac{|Top_N(\mathbb{G})\cap Top_N(\mathbb{I})|}{N},
\end{displaymath}
which refers to the percentage of locations that belongs to both of
Top $N$ locations in reported mobility traces and inferred mobility
traces.

\begin{figure}[htpb]
  \centering
  \includegraphics[width=0.32\textwidth]{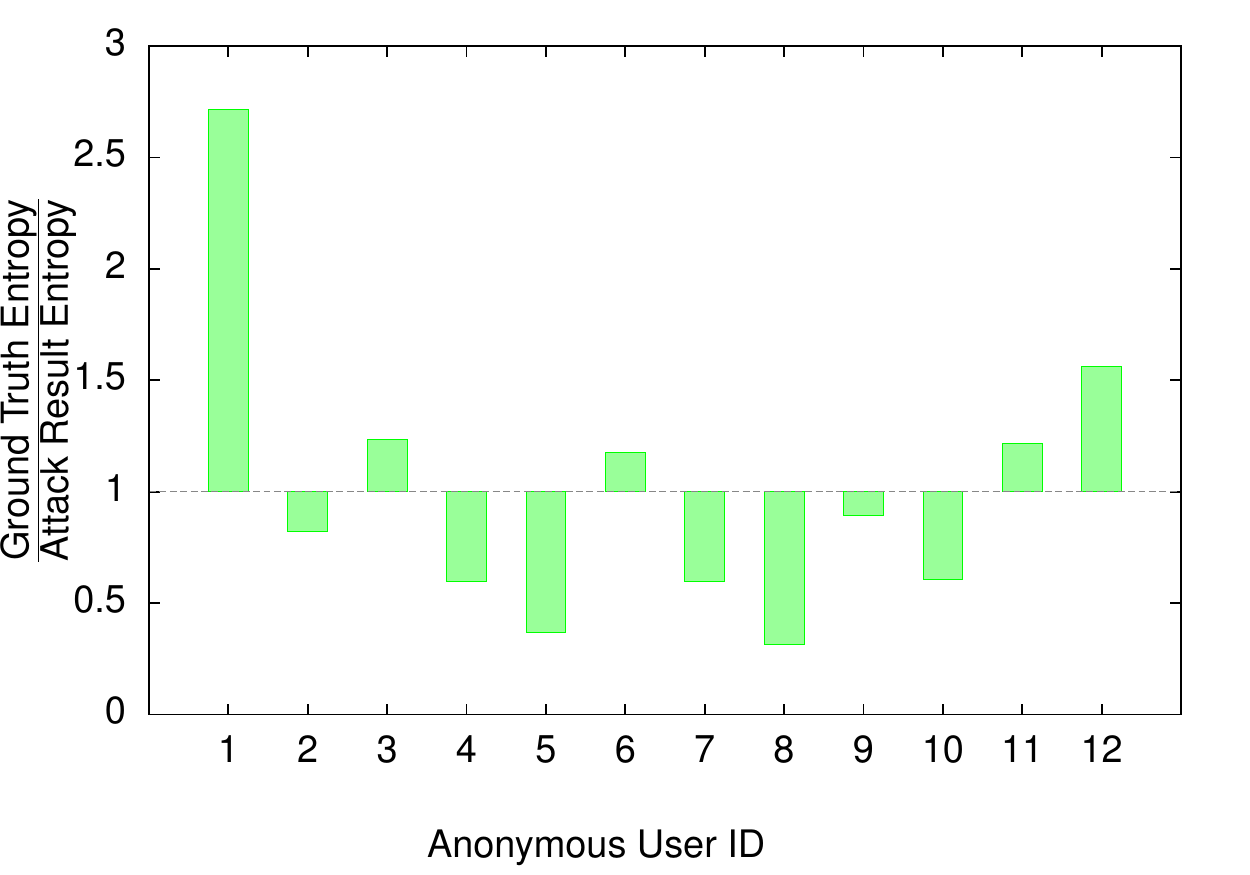}
  \caption{Ground Truth vs. Attack Result in Entropy}
  \label{fig:entropy}
\end{figure}

\begin{table*}
  \centering
  \begin{tabular}{|c|c c c|c c c|c c c|}
    \hline
    \multirow{2}{*}{top location} & \multicolumn{3}{|c|}{one week} & \multicolumn{3}{|c|}{two weeks} & \multicolumn{3}{|c|}{three weeks}\\
    \cline{2-10}
    & Momo & Wechat & Skout & Momo & Wechat & Skout & Momo & Wechat & Skout\\
    \hline
    $1$ & $92.3\%$ & $50.0\%$ & $20.0\%$ & $100.0\%$ & $57.1\%$ & $60.0\%$ & $100.0\%$ & $71.4\%$ & $60.0\%$ \\
    $2$ & $46.1\%$ & $21.4\%$ & $0.0\%$ & $46.1\%$ & $21.4\%$ & $40.0\%$ & $69.2\%$ & $21.4\%$ & $40.0\%$ \\
    $3$ & $30.7\%$ & $21.4\%$ & $20.0\%$ & $46.1\%$ & $28.5\%$ & $60.0\%$ & $38.4\%$ & $28.5\%$ & $80.0\%$ \\
    $4$ & $23.0\%$ & $35.7\%$ & $20.0\%$ & $30.7\%$ & $35.7\%$ & $40.0\%$ & $38.4\%$ & $35.7\%$ & $40.0\%$ \\
    $5$ & $23.0\%$ & $21.4\%$ & $0.0\%$ & $15.3\%$ & $21.4\%$ & $40.0\%$ & $15.3\%$ & $14.2\%$ & $40.0\%$ \\
    \hline
  \end{tabular}
  \caption{Top 5 Location Coverage Result for 3 Weeks}
  \label{table:toploc}
\end{table*}

\textbf{Impact of Usage Pattern:} Interestingly, our experiment
results show that the inferred Top N locations of users do not exactly
match to their real Top N locations. For example, for some users, the
inferred Top 1 location may be the Top 2, or 3 location in their real
traces just as shown in Table \ref{table:toploc} that Top 2 location
coverage sometimes is less than Top 3 coverage. In our experiments, it
is found that only $65\%$ volunteers' top $1$ locations exactly match
with their top $1$ locations in ground truth trace. This can be
explained by investigating the difference between every user's real
traces and his usage pattern.  We adopt the definition of location
entropy\cite{24-shokri2011quantifying}, which is defined as:
\begin{equation}
  \label{eq:entropy}
  H(x)=-\sum_{x\in Loc} p_{x} \mbox{log} p_{x}
\end{equation}
where \emph{Loc} is the location set consisting of the locations that
a user visited, $p_x$ is the probability that the user is at the
location $x$. From this definition, it is observed that the bigger the
location entropy is, the more diversified the user's locations are.
Here, we measure both of the ground truth trace location entropy $H_1$
and the inferred location entropy $H_2$.  $H_1$ indicates how many
places the users visited, while $H_2$ shows how many places the users
use the app. Therefore, the metric $H_1/H_2$ shows the usage pattern,
which is shown in Fig \ref{fig:entropy}.  In Fig \ref{fig:entropy}, we
randomly sample $12$ volunteers' ground truth location to tracking
location rate. Fig \ref{fig:entropy} shows that the ground truth trace
location entropy $H_1$ is not always close to the inferred location
entropy $H_2$. For example, a user spends most of his time at his home
and office which makes the value $H_1$ very small, but he may always
uses the LBSN app in different places which will induce a large $H_2$
and a small $H_1/H_2$ rate. Similarly, if a user travels to many places
which leads to a large $H_1$, but he mainly uses at his home so that
his $H_2$ is very small (i.e., close to 0) and the rate will be
large. Therefore, from the attacker point of view, Top $N$ locations
should be regarded of the same importance, which motivates us to use
unordered Top $N$ locations in our evaluation.

\textbf{Evaluation Results:} Without loss of the generality, we set
$N=5$ and evaluate the top location coverage rate for three weeks,
which is shown in Table \ref{table:toploc}. From Table
\ref{table:toploc}, it is observed that Momo shows the best coverage
rate. After three weeks tracking, we can obtain all the volunteers'
top $1$ locations and about $70\%$ volunteers' top $2$ locations. For
Wechat, we could successfully infer $71.4\%$ $21.4\%$, $28.5\%$ of top
$1, 2, 3$ locations after $3$ week tracking. For Skout, $60.0\%$,
$40.0\%$, $80.0\%$ volunteers's top $1, 2, 3$ locations could be
successfully recovered. Our evaluation results also show that the
temporal factor plays an important role in Top $N$ location
recovery. In particular, the Top $N$ location coverage rate will
significantly increase along with more tracking days. In general,
FreeTack shows a high Top $5$ location coverage rate.



\section{Attacks Mitigation}
\label{sec:mitigation}
In this section, we aim to propose some suggestions to limit the
attacking capability of the attackers. We hope that the following
discussion would raise location privacy awareness of the LBSN
developers and would also inspire other researchers to find more
advanced protection techniques.

\subsection{Prevention of Using LBSN as Location Oracle}

Firstly, the feasibility of considered attack roots in the fact that
these app servers retrieve location without requiring effective
location proofs. One potential countermeasure for the considered
attack is adding some location proof modules to ensure the
authenticity of the locations. The typical location proof techniques
include using the deployed trusted infrastructures (e.g., cell tower
or Wi-Fi access points) \cite{saroiu2009enabling} or using
environmental signals as the location tags
\cite{17-narayanan2011location, 16-zheng2012sharp,
  brassil2013traffic}. However, they either require the presence of
trusted infrastructure or are only effective in a small-scale (e.g.,
less than 100m) due to the spatial diversity of wireless
signals. As a result, the existing location proof techniques are only
feasible in a small-scale region and less practical in our scenario.

To limit the attacker's capability, the service provider could compare
users' location changes with their mobility patterns or behavior
patterns to identify the potential anomalous users (e.g, changing the
locations too frequently or making too many queries within a short
period). For example, from our experiment, it is observed that Wechat
has put a limit on the number of queries issued at a certain duration
(depending on the workload of the server) and the misbehaving account
will be blocked for a specific period, which significantly slows down
attacking process. Our real-world experiments show that, though the
attacker may use multiple accounts to speed up the attack, a more
stringent limit on the number of queries will increase the difficulty
of launching the attacks since the attack should be finished between
two consequent location updating events of the target.

\subsection{A User Controllable Privacy Enhancement Framework}

In this subsection, we present a user controllable privacy enhancement
framework. Our basic idea is that we could use a global grid reference
system to generate the relative distance, providing obfuscation
functionalities to mobile users. Even though some advanced obfuscation
techniques including enlarging the radius of cloaking region or
shifting the location by randomly generated distance and rotation
angle can be implemented at the client side
\cite{ardagna2011obfuscation}, the major limitation for client-based
obfuscation technique is that different users may have different
privacy protection levels and acceptable utility levels (the measurement
error of relative distance). However, the relative distance error is
determined by the geo-location error of both parties, which may exceed
the acceptable utility level of any single party.

\begin{figure}[htbp]
  \centering
  \begin{subfigure}[htbp!]{0.23\textwidth}
    \centering
    \includegraphics[width=\textwidth]{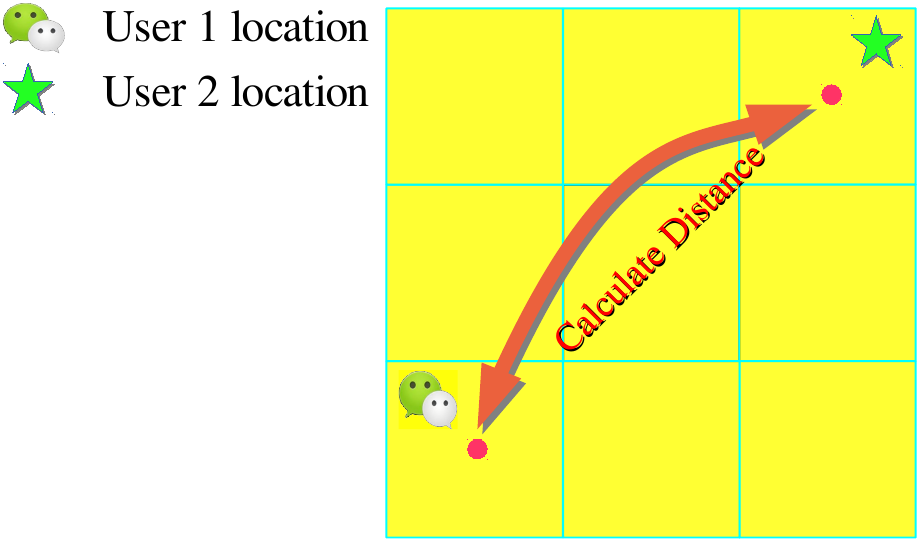}
    \caption{Basic Grid Reference System}
    \label{fig:grs}
  \end{subfigure}
  \begin{subfigure}[htbp!]{0.180\textwidth}
    \centering
    \includegraphics[width=\textwidth]{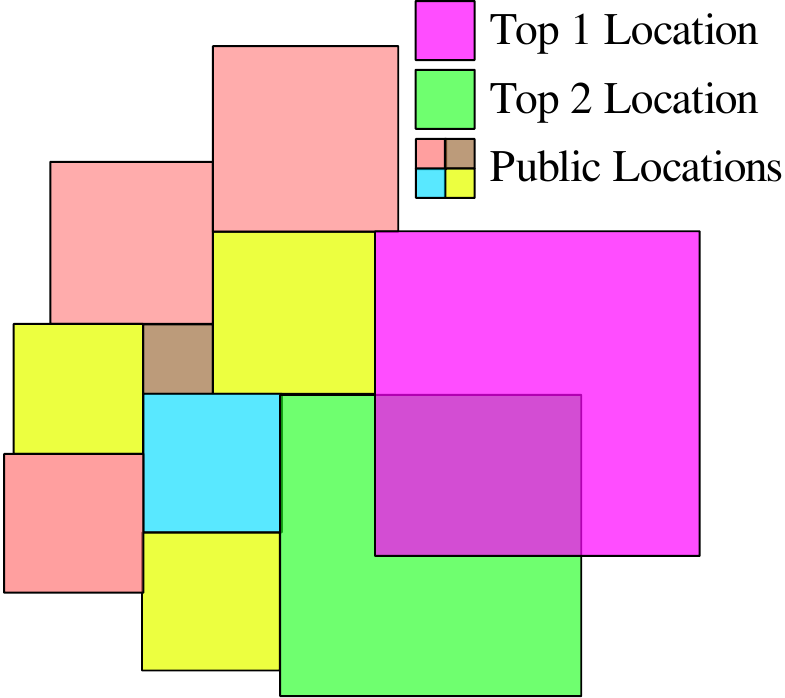}
    \caption{Classified Grid Reference System}
    \label{fig:cgrs}
  \end{subfigure}
  \caption{The Grid Reference System}
  \label{fig:grs-protection}
\end{figure}

\begin{figure}[htbp]
  \centering
  \includegraphics[width=0.32\textwidth]{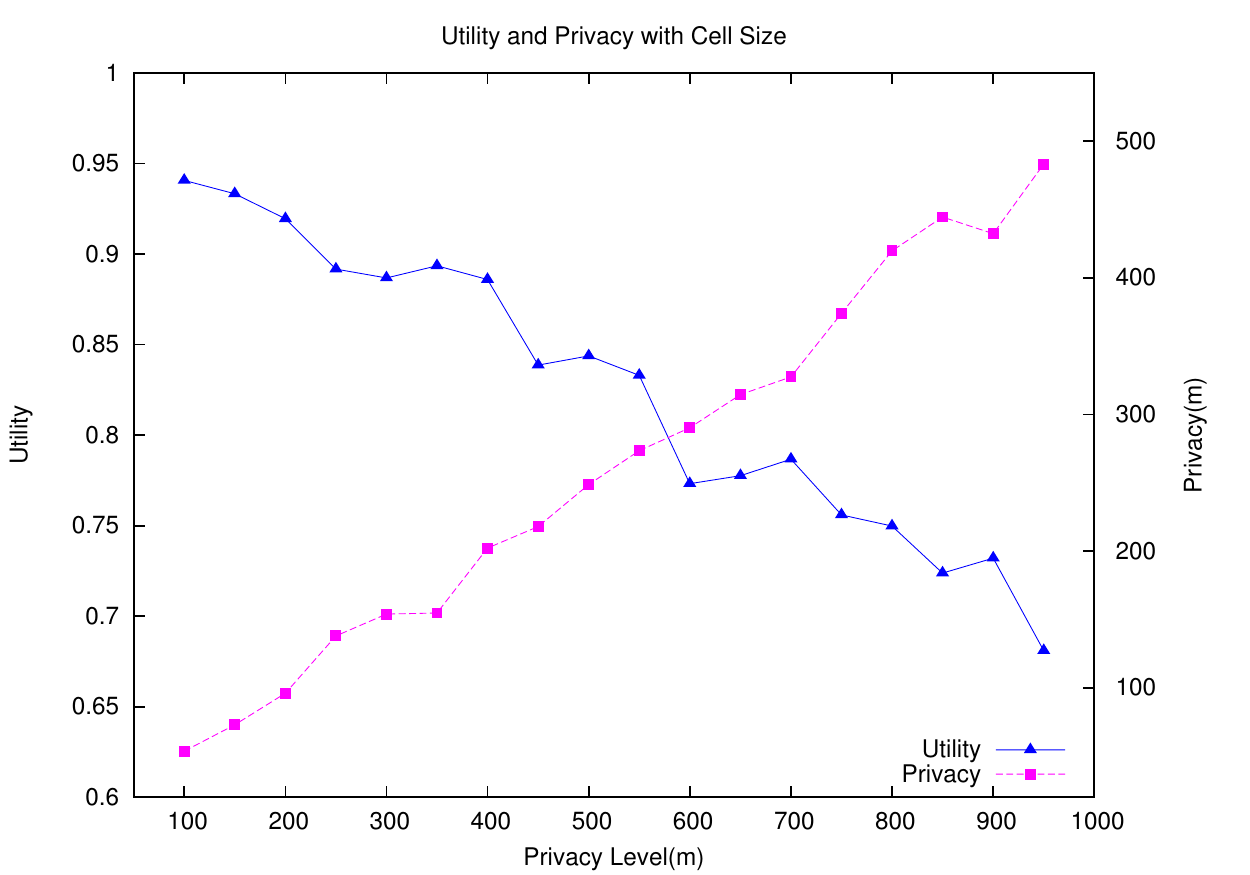}
  \caption{Relationship of Utility/Privacy with Cell Size}
  \label{fig:upc}
\end{figure}

\textbf{Distance Obfuscation with Grid Reference System:} In this
work, we propose a distance obfuscation based on grid reference
system, which aims to prevent the attacker from using LBSN as the
location oracle to obtain the accurate location information. As shown
in Fig \ref{fig:grs}, the server maintains a grid reference system,
where the location of a mobile user can be expressed as the center of
the grid cell that the user is located in. Therefore, the relative
distance of two users is expressed by the distance of two grid cells
defined as the minimum path connecting these two cells. The benefit of
using grid reference system to express the relative distance of two
users is that it obfuscates the real location of mobile users with the
center of the cell and the attacker cannot obtain extra information of
the target if the generated fake anchors are located at the same
cell. Similar to other obfuscation techniques, grid reference system
will also decrease user utility. Considering the relative distance is
the main metric of LBSN, we define the metric of privacy as:
\begin{displaymath}
  Privacy=Dist(Loc_{R}, Loc_{O}),
\end{displaymath}
where $Loc_{R}$ and $Loc_{O}$ refer to the real and obfuscated
location of the mobile user, respectively, and function $Dist()$
returns the distance of two locations in Grid reference system. By
given a specific anchor node at location $Loc_A$, we further define
the utility metric as
\begin{displaymath}
  Utility=1-\frac{|DDist(Loc_{R}, Loc_A) - DDist(Loc_{O}, Loc_A)|}{Dist_{max}},
\end{displaymath}
where function $DDist()$ returns the displayed distance in LBSN apps,
$Dist_{max}$ represents the maximum distance that the user could
tolerate. It is obvious that, when the displayed distance between the
real location and anchor point $DDist(Loc_{R}, Loc_A)$ equals
displayed distance between the obfuscated location and anchor point
$DDist(Loc_{O}, Loc_A)$, the utility achieves the maximum value
$1$. When $DDist(Loc_{O}, Loc_A)$ is much larger or smaller than
$DDist(Loc_{R}, Loc_A)$ (their gap should not be larger than
$Dist_{max}$), the utility is close to $0$. By assigning different
values to the size of the cells, we could achieve different location
privacy protection level as well as different utilities. We evaluate
the effectiveness of location privacy protection and its impact on the
utility by applying grid reference system to the data set collected
from our real world experiments (ground truth data and inferred
location data). Fig \ref{fig:upc} shows the privacy gain and the
utility under different settings of cell size. It is observed that the
increase of privacy gain will lead to the decrease of the utility, and
vice versa. We will discuss how to achieve the
tradeoff of privacy and utility next.

\begin{figure}[htbp]
  \centering
  \includegraphics[width=0.32\textwidth]{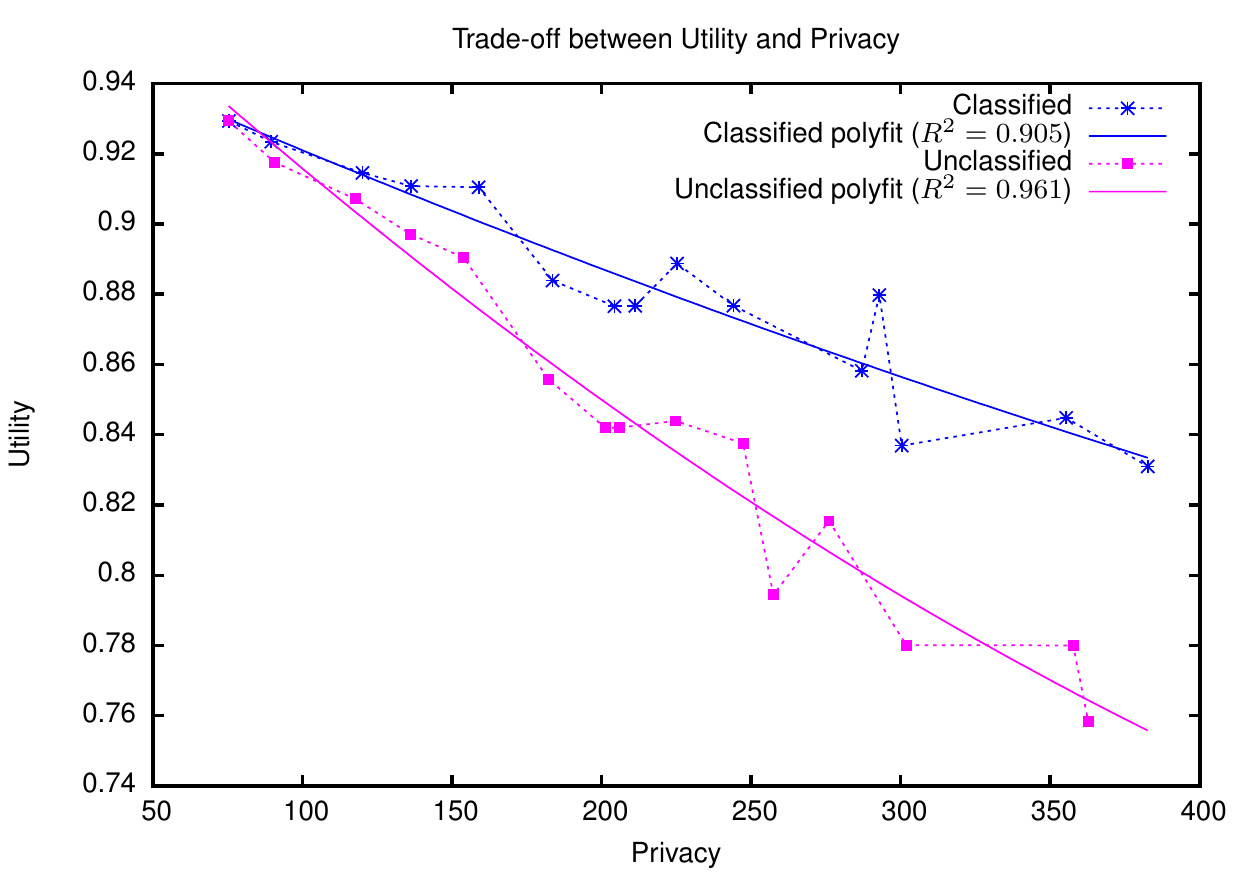}
  \caption{Comparison of Utility/Privacy Trade-offs}
  \label{fig:tupc}
\end{figure}

\textbf{Achieving Privacy and Utility Trade-off via Classification of
  Users' Locations:} In the previous section, we have shown that
the obfuscation techniques will bring the decrease of the utility. To
achieve the tradeoff between the privacy and utility, we introduce a
novel user controllable location privacy protection scheme. The
proposed scheme is motivated from the observation that the
user has different location privacy protection preference for
different locations. For example, a mobile user cares more about their
Top $2$ location privacy (e.g., home, work place) while care less
about the location privacy issue when he is at public regions (e.g.,
cafe or bars), which makes him have different privacy protection
requirements. Therefore, in a user controllable location privacy
protection solution, the mobile users could classify the locations
into several categories, which correspond to different privacy
protection requirements as well as different obfuscation
parameters. During the subsequent LBSN usage process, users
record their location profiles that are ranked with their visiting
frequency and could be dynamically updated along with users'
usage. With such a location profile with different ranking, the most
frequently visited locations are given more privacy protection and
thus suffer from a lower utility while the less frequently visited
locations could enjoy more utility with less privacy protection as
indicated in Fig \ref{fig:cgrs}. To implement our idea, we transform
the original grid reference system of the uniform cell size to the
non-uniform grid reference system, in which top locations cover a
larger area while public regions cover a smaller area. Note that the
proposed location classification concept could also be applied to
other existing obfuscation techniques \cite{ardagna2011obfuscation}.
To evaluate the proposed solution, we compare the uniform grid
reference system with non-uniform grid reference system based on the
data set collected from our real-world experiments. In the uniform
grid reference system, we tune the cell size from 200m to 1000m,
which correspond to the privacy level from 50 to 400. In the
non-uniform grid reference system, we fix the cell size of top
locations to 1000m to provide highest privacy protection level while
tune the cell size of normal location from 200m to 1000m. It is
observed that the non-uniform grid reference system based on location
classification has a significant advantage in privacy/utility
trade-off over the uniform grid
reference system as shown in Fig \ref{fig:tupc}.

We notice that Momo and Wechat provide an option to manually remove
their locations from the public. However, with no idea about the
potential risks brought by LBSN apps, few people do choose this
option. This further signifies the importance of making the public
more aware of the potential risk, which is one of major motivations of
this paper.


\section{Related Work}
\label{sec:related}
Location Privacy Protection in location-based services is a
long-standing topic and has received a lot of attentions in the last
decades. The most popular approach to achieve location privacy in LBS
is utilizing the obfuscation techniques to coarse the spatial or
temporal granularity of the users' real locations
\cite{rel1-gruteser2003anonymous,rel2-hoh2007preserving}.  A different
approach to hide the users' location is based on mix zones.  Mix zones
are defined as the regions where users keep silent while changing
their pseudonyms together\cite{20-freudiger2009optimal}.  The third
approach is to protect location privacy by adding dummy requests,
which are issued by fake location and indistinguishable from real
requests \cite{23-xu2009feeling}. A recent work
\cite{22-shokri2012protecting} proposes a game-theoretic framework
that enables a designer to find the optimal LPPM for a given
location-based service, ensuring a satisfactory service quality for
the user. As shown in the paper that, it is possible to apply various
obfuscation techniques to enhance the location privacy of
LBSN. Different from the location privacy issues considered in
previous works, providing the relative distance is the key
functionality of LBSN apps while the obfuscation will inevitably
reduce the utility of LBSNs. Therefore, how to achieve the tradeoff
between the location privacy and the utility is of the highest
priority. The proposed users' location classification based approach
could help to reduce the impact of obfuscation techniques on users'
utility, and thus can be a compliment to various obfuscation
techniques.


There are many other works which study how to infer the victim's
trajectory and further re-identify his other private information
\cite{golle2009anonymity, 12-zang2011anonymization,
  11-srivatsa2012deanonymizing, 9-de2013unique}. This work is
different from the existing work in that we propose an novel attack
approach, which could be exploited by anyone to perform an involuntary
tracking towards any specific target. The collected tracking traces
could be used for user re-identification.

There are some research efforts targeting at secure friend discovery
in mobile social networks\cite{dong2011secure, li2011findu,
  zhang2012fine}. These works consider
testing equalility between attributes in profiles and setting threshold on
number of matching pairs, which is different from our considered problem.

We also notice that there are some smart-phone privacy leaking work,
most of which focus on various types of mobile malware on various
platforms of iOS, Android and Symbian\cite{felt2011android,
  reynaud2012freemarket, jiang2013detecting}. Our work is different
from existing works in that the proposed attack is actually based on
one of system design drawbacks. To the best of our knowledge, our work
is the first one to investigate the location privacy leaking issue
from LBSN apps.


\section{Conclusion}
\label{sec:conclusion}
LBSN is becoming extremely popular recently. However, most LBSN users
are unaware of the location privacy leaking issue.  We target at 3
most popular LBSN apps and develop a novel automatic tracking system,
which could achieve range-free, accurate, and involuntary tracking
towards the target by only using the public information. Our
real-world attack experiments show that the attack could achieve high
localization accuracy and the attacker could recover the users' top 5
locations with high possibility and hence, in addition to the malware,
inappropriate location privacy protection techniques of LBSNs pose a
more serious threat in practice. We've discussed various mechanisms to
mitigate such threats and analysed the privacy and utility
trade-off. As the first work of its kind, our study is expected to
urge LBSN service providers to revisit their location privacy
protection techniques and call for more attentions from the public to
have the full knowledge of the potential risks brought by LBSN
apps. Our mitigation suggestions will provide a guideline for future
revisions of these LBSN apps.

\bibliographystyle{IEEEtran}
\bibliography{muyuan2013all}

\end{document}